\documentclass[%
 preprintnumbers,
 notitlepage,
 nofootinbib,
 nobibnotes,
 eqsecnum,
 amsmath,amssymb,
 aps,
]{revtex4-1}

\usepackage[dvipdfmx]{graphicx,color}
\usepackage{dcolumn}
\usepackage{bm}
\usepackage{hyperref}
\usepackage{xcolor}
\hypersetup{
    colorlinks=true,
    citecolor=blue,
    linkcolor=magenta,
    urlcolor=teal,
}
\usepackage{mathtools}

\begin{document}

\preprint{KOBE-COSMO-20-08}

\title{Stability of magnetic black holes in general nonlinear electrodynamics}

\author{Kimihiro Nomura}%
 \email{190s111s@stu.kobe-u.ac.jp}
 \affiliation{%
Department of Physics, Kobe University, Kobe 657-8501, Japan
}%
\author{Daisuke Yoshida}
 \email{dyoshida@hawk.kobe-u.ac.jp}
 \affiliation{%
Department of Physics, Kobe University, Kobe 657-8501, Japan
}%
\author{Jiro Soda}%
 \email{jiro@phys.sci.kobe-u.ac.jp}
 \affiliation{%
Department of Physics, Kobe University, Kobe 657-8501, Japan
}%

%
%

\date{\today}

\begin{abstract}
We study the perturbative stability of magnetic black holes in a general class of nonlinear electrodynamics, where the Lagrangian is given by a general function of the field strength of electromagnetic field $F_{\mu\nu}$ and its Hodge dual $\widetilde{F}_{\mu\nu}$. 
We derive sufficient conditions for the stability of the black holes.
We apply the stability conditions to 
Bardeen's regular black holes, black holes in Euler--Heisenberg theory, and black holes in Born--Infeld theory.
As a result, we obtain a sufficient condition for the stability of  Bardeen's black holes, 
which restricts $F_{\mu\nu}\widetilde{F}^{\mu\nu}$ dependence of the Lagrangian. We also show that
black holes in Euler--Heisenberg theory are stable for a sufficiently small magnetic charge.  
Moreover, we prove the stability of black holes in the Born--Infeld electrodynamics 
even when including $F_{\mu\nu}\widetilde{F}^{\mu\nu}$ dependence. 
\end{abstract}

                              
\maketitle

\section{Introduction}
The existence of black holes in the Universe has been confirmed by direct detection of gravitational waves from a binary \cite{Abbott:2016blz} and imaging of a black hole shadow \cite{Akiyama:2019cqa}.
There is no doubt that the black holes have gained more importance not only in astrophysics but also in theoretical physics.
In particular, they provide us with profound understanding of gravity.
A primary concern about black holes is the perturbative stability, which has been mathematically challenging.
The simplest class of black holes found in general relativity  are proved to be stable.
Historically, the linear perturbation theory around Schwarzschild black holes was first developed by Regge, Wheeler \cite{Regge:1957td} and Zerilli \cite{Zerilli:1970se}, and then it was extended to Reissner--Nordstr\"om black holes in Einstein--Maxwell system in Refs.\,\cite{Zerilli:1974ai,Moncrief:1974gw,Moncrief:1974ng,Moncrief:1975sb}. 

Curiously, however, there exists a singularity inside a black hole.
Indeed, the singularity theorem \cite{Penrose:1964wq, Hawking:1969sw,Hawking:1973uf,Wald:1984rg} states that a formation of a singularity is inevitable in classical general relativity with a matter which satisfies energy conditions. 
It is  generally believed that the singularity is resolved by quantum effects.
From this point of view,  general relativity is merely a low energy effective theory
 of ultraviolet complete quantum theory of gravity. 
 Hence, we should also incorporate quantum corrections of matter. In the electromagnetic case, for example, 
 we should consider Euler--Heisenberg theory \cite{Heisenberg:1935qt}
 which include $\widetilde{{\cal F}}\propto  F_{\mu\nu}\widetilde{F}^{\mu\nu}$ dependence, where $F_{\mu\nu}$ and $\widetilde{F}_{\mu\nu}$ are the field strength of electromagnetic field  and its Hodge dual.
An attempt to resolve the singularity may affect the stability of black holes. Thus, it is important to analyze the stability of black holes with the quantum corrections.

Note that a curvature singularity accompanies the divergence of curvature invariants, which indicates the existence of a cutoff scale which divide between classical general relativity and fundamental theory. The similar issue has been considered in electrodynamics and the idea of limiting field strength was invented by Born and Infeld,  which is called Born--Infeld theory of electrodynamics \cite{Born:1934gh}.  Since the Born--Infeld theory has an overlap with Euler--Heisenberg theory,
it is also worth examining the stability of Born--Infeld black holes.
Intriguingly, the idea of Born and Infeld can be promoted to the idea of ``limiting curvature hypothesis'' \cite{markov1982limiting}, which states that there is an upper bound on curvature invariants in fundamental theory. The limiting curvature hypothesis suggests that the singularity of a black hole should be replaced with a regular structure \cite{Frolov:1989pf,Frolov:1988vj, Barrabes:1995nk}. A dynamical model trying to realize limiting curvature hypothesis called ``limiting curvature theory'' was studied initially in the context of cosmology \cite{Mukhanov:1991zn,Brandenberger:1993ef,Moessner:1994jm,Easson:2006jd,Yoshida:2017swb,Yoshida:2018ndv}.
Then the theory is applied to black hole singularity in Ref.\,\cite{Yoshida:2018kwy}, though the singularity cannot be removed there.
A realization of the limiting curvature hypothesis in the context of scalar-tensor theories is also studied in Refs.\,\cite{Chamseddine:2016uef,Chamseddine:2016ktu,Quintin:2019orx}.
Unfortunately, the regular black hole in Ref.\,\cite{Chamseddine:2016ktu} seems to be unstable \cite{Takahashi:2017pje, Ijjas:2016pad, Firouzjahi:2017txv}.

To our best knowledge, the first model of a regular black hole was proposed by Bardeen \cite{bardeen1968non}. Note that the regular black hole metric
was obtained without assuming any specific theory. 
Remarkably, Ay\'on-Beato and Garcia found that a class of  regular black holes can be obtained as a solution of general relativity with nonlinear electrodynamics \cite{AyonBeato:1998ub}. 
Moreover, the same authors found that the Bardeen black hole can be also obtained as a solution of nonlinear electrodynamics with a magnetic charge \cite{AyonBeato:2000zs}. The Bardeen black hole solution can be obtained also with an electric charge \cite{Rodrigues:2018bdc}. Other black hole solutions in nonlinear electrodynamics have been studied in Refs.\,\cite{AyonBeato:1999rg,AyonBeato:1999ec,Bronnikov:2000vy,AyonBeato:2004ih, Stefanov:2007qw, Hassaine:2008pw,Uchikata:2012zs,Balart:2014cga, Fernando:2016ksb, Toshmatov:2017zpr, Kruglov:2017mpj, Bronnikov:2017sgg, Ali:2018boy, Li:2018bny, Poshteh:2020sgp, Cisterna:2020rkc}. 

Given that there are regular black holes in nonlinear electrodynamics, it is worth studying if the resolution of a singularity affects the stability of black holes in  nonlinear electrodynamics.  Indeed, the analysis of the linear perturbations was extended to a class of nonlinear electrodynamics in  Ref.\,\cite{Moreno:2002gg}, where the Lagrangian of electromagnetic field is assumed to be a function of ${\cal F}\propto F_{\mu\nu}F^{\mu\nu}$.
Then, stability analysis, including thermodynamical stability, of this class of nonlinear electrodynamics was also studied in Refs.\,\cite{Breton:2005ye, Breton:2007bza, Man:2013hpa, Breton:2014nba, Toshmatov:2019gxg, Sharif:2020bzt}. The quasi-normal modes of black holes in nonlinear electrodynamics were studied in Refs.\,\cite{Stefanov:2009zza,Doneva:2010ke, Fernando:2012yw, Chaverra:2016ttw, Xi:2016qrg, Wu:2017tfo, Wu:2018xza,Toshmatov:2018tyo,Toshmatov:2018ell, Dey:2018cws, Panotopoulos:2019qjk, Flachi:2012nv}.
 
 It should be emphasized that $\widetilde{{\cal F}}$ dependent models have not been considered in the above analysis.
 Since quantum corrections include $\widetilde{{\cal F}}$ dependence, it is legitimate to expect that there are  other  regular black hole solutions
  in general nonlinear electrodynamics. Thus,  
 it is worth studying the stability of black holes in general nonlinear electrodynamics where a Lagrangian 
   depends on both ${\cal F}$ and $\widetilde{{\cal F}}$.
In the present paper, we achieve this aim by extending the analysis of Ref.\,\cite{Moreno:2002gg} to general nonlinear electrodynamics. 
 Our analysis includes the stability analysis of black holes in Born--Infeld theory and Euler--Heisenberg theory. Note that a consistency condition from causality and unitarity of this class of theory is studied in Ref.\,\cite{Shabad:2011hf}.

This paper is organized as follows.
In Section \ref{sec:background}, we investigate  spherically symmetric black hole solutions with a magnetic charge in general relativity coupled to nonlinear electrodynamics.
In Section \ref{sec:perturbations}, we derive master equations of motion for metric and electromagnetic perturbations in the black hole  background.
In Section \ref{sec:stability}, we study the sufficient conditions for the stability of  the magnetic black holes. 
In Section \ref{sec:applications}, we perform the stability analysis for specific models in nonlinear electrodynamics as applications. 
The Section \ref{sec:discussion} is devoted to the summary of the results.
In Appendix \ref{app:General form}, we study a general form of Lagrangian in nonlinear electrodynamics.
In Appendix \ref{app:gauge}, the expansions of the variables with a basis of spherical harmonics and gauge transformations for metric and electromagnetic perturbations are carried out explicitly. 

Throughout this paper, we use the conventions such that the metric signature is $(-,+,+,+)$ and the covariant antisymmetric Levi-Civita tensor is normalized as $\epsilon_{0123} = \sqrt{-g}$, where $g$ is the determinant of the metric.
We work in a unit system where the speed of light $c$, the reduced Planck constant $\hbar$, and the permittivity in a vacuum $\varepsilon_{0}$ are all equal to one.

\section{Black holes with a magnetic charge}
\label{sec:background}

We consider the action in general relativity with a cosmological constant coupled to general nonlinear electrodynamics, which contains the contributions not only from a field strength $\bm{F} = \frac{1}{2} F_{\mu\nu} dx^{\mu} \wedge dx^{\nu} = d\bm{A}$ but also from its Hodge dual $\widetilde{\bm{F}} = \frac{1}{2} \widetilde{F}_{\mu\nu} dx^{\mu} \wedge dx^{\nu} = \frac{1}{4} \epsilon_{\mu\nu\rho\sigma} F^{\rho\sigma} dx^{\mu} \wedge dx^{\nu}$, where $\bm{A} = A_{\mu} dx^{\mu}$ is an electromagnetic field.
The action is given by
\begin{align}
S[\bm{g}, \bm{A}] = \int d^{4}x \sqrt{-g}
\left[
\frac{1}{16 \pi G}(R - 2\Lambda) -  \mathcal{L}\left(\mathcal{X}_{n_{1}, m_{1}, n_{2}, m_{2}, \dots}
\right)
\right] \, ,
\label{eq2-0-1}
\end{align}
where $R$ represents the Ricci scalar with respect to the spacetime metric tensor $\bm{g} = g_{\mu\nu} dx^{\mu} dx^{\nu}$, $\Lambda$ is the cosmological constant, $G$ is the Newton's gravitational constant, $g$ is the determinant of $g_{\mu\nu}$, and $\mathcal{L}$ is an arbitrary function of scalars defined by
\begin{align}
\mathcal{X}_{n_{1}, m_{1}, n_{2}, m_{2}, \dots}
\coloneqq
\mathrm{tr}(
F^{n_{1}} \widetilde{F}^{m_{1}} F^{n_{2}} \widetilde{F}^{m_{2}} \cdots
)
\, 
\label{eq2-0-2}
\end{align}
with non-negative integers $n_{1}, m_{1}, n_{2}, m_{2}, \dots$, where $F$ and $\widetilde{F}$ stand for the matrices with components ${F_{\mu}}^{\nu}$ and ${\widetilde{F}_{\mu}}^{~\nu}$, respectively.
The quantity \eqref{eq2-0-2} represents an arbitrary scalar constructed from the field strength and its Hodge dual.
For example,
\begin{align}
&\mathcal{X}_{2,0} = \mathrm{tr} (F^{2}) = {F_{\mu}}^{\nu} {F_{\nu}}^{\mu} = -F_{\mu\nu} F^{\mu\nu} \, ,
\label{eq2-0-3} \\
&\mathcal{X}_{1,1} = \mathrm{tr} (F \widetilde{F}) = {F_{\mu}}^{\nu} {\widetilde{F}_{\nu}}^{~\mu}
= -F_{\mu\nu} \widetilde{F}^{\mu\nu} \, .
\label{eq2-0-4}
\end{align}
When we choose $\mathcal{L} = - \frac{1}{4} \mathcal{X}_{2,0} = \frac{1}{4} F_{\mu\nu} F^{\mu\nu}$, the action \eqref{eq2-0-1} gives the familiar Einstein--Maxwell theory with a cosmological constant.

As we discuss in Appendix \ref{app:General form}, the general action given by Eq.\,\eqref{eq2-0-1} eventually reduces to the following form,
\begin{align}
S[\bm{g}, \bm{A}] = \int d^{4}x \sqrt{-g}
\left[
\frac{1}{16 \pi G}(R - 2\Lambda) -  \mathcal{L}(\mathcal{F}, \widetilde{\mathcal{F}})
\right],
\label{eq2-1}
\end{align}
where  $\mathcal{L}$ is an arbitrary function of the invariants defined by
\begin{align}
\mathcal{F} &\coloneqq \frac{1}{4} F_{\mu\nu} F^{\mu\nu} \, ,
\label{eq2-1-2}\\
\widetilde{\mathcal{F}} &\coloneqq \frac{1}{4} F_{\mu\nu} \widetilde{F}^{\mu\nu} = \frac{1}{8} \epsilon_{\mu\nu\rho\sigma} F^{\mu\nu} F^{\rho\sigma} .
\label{eq2-1-3}
\end{align}
By taking variation of the action \eqref{eq2-1} with respect to the metric, we obtain the Einstein equations
\begin{align}
&{G_{\mu}}^{\nu} = 8 \pi G {T_{\mu}}^{\nu} \,,
\label{eq2-2}
\end{align}
where ${G_{\mu}}^{\nu}$ stand for the components of the Einstein tensor calculated from $g_{\mu\nu}$, and ${T_{\mu}}^{\nu}$ are the components of the energy-momentum tensor of the electromagnetic field,
\begin{align}
{T_{\mu}}^{\nu} 
= \mathcal{L}_{\mathcal{F}} F_{\mu\lambda} {F}^{\nu\lambda} 
+ \delta_{\mu}^{\nu} \left( \mathcal{L}_{\widetilde{\mathcal{F}}} \widetilde{\mathcal{F}} - \mathcal{L} - \frac{\Lambda}{8 \pi G} \right) \,,
\label{eq2-2-2}
\end{align}
where $\mathcal{L}_{\mathcal{F}} \coloneqq \partial \mathcal{L}/\partial {\mathcal{F}}$, $\mathcal{L}_{\widetilde{\mathcal{F}}} \coloneqq \partial \mathcal{L}/\partial{\widetilde{\mathcal{F}}}$.
Similarly, the equations of motion for the electromagnetic field derived by varying the action \eqref{eq2-1} with respect to $A_{\mu}$ are given by
\begin{align}
\nabla_{\mu} \left( \mathcal{L}_{\mathcal{F}} F^{\mu\nu} + \mathcal{L}_{\widetilde{\mathcal{F}}} \widetilde{F}^{\mu\nu} \right) = 0 \,,
\label{eq2-3}
\end{align}
where $\nabla_{\mu}$ stand for the covariant derivative with respect to $g_{\mu\nu}$.

{
In this paper, we focus on black holes with a magnetic charge because the analysis for magnetic black holes can be translated  into that for electrically charged black holes through the electromagnetic duality of nonlinear electrodynamics \cite{Gibbons:1995cv}.
Thus, let us seek the static and spherically symmetric solution with a magnetic charge below.
}
In this case, we can put an ansatz for the metric as
\begin{align}
\bm{g} 
= - f(r) \, dt^{2} + h(r) \, dr^{2} + r^{2} \, d\Omega^{2}\,,
\label{eq2-4}
\end{align}
where $f(r)$ and $h(r)$ are some functions of the radial coordinate $r$, and $d\Omega^{2} = d\theta^{2} + \sin^{2} \theta  \, d\phi^{2}$ represents the line element on the unit two-sphere $S^{2}$.
Now we are interested in the solutions with a static and spherically symmetric magnetic field, thus we take the configuration of the background field strength as
\begin{align}
\bm{F} = q \sin \theta \, d\theta \wedge d\phi  \ .
\label{eq2-5}
\end{align}
Here, $q$ must be a constant because of the Bianchi identity, $d \bm{F} = 0$.
Then we have the background values of the invariants as
\begin{align}
\mathcal{F} = \frac{q^{2}}{2r^{4}}\,, \qquad
\mathcal{\widetilde{F}} = 0 \,.
\label{eq2-10}
\end{align}
The quantity $q$ is interpreted as a magnetic charge in this spacetime,
\begin{align}
\int_{S^{2}} \bm{F}
= \int_{0}^{\pi} d \theta \int_{0}^{2\pi} d\phi \, \sin \theta \, q
= 4 \pi q \, ,
\label{eq2-10-2}
\end{align}
where the left-most term implies that the integral is taken over $S^{2}$.
Note that we can take the background gauge potential which induces the field strength \eqref{eq2-5} as
\begin{align}
\bm{A} = \bm{A}_{\pm} = q (\pm 1 - \cos \theta) \, d\phi \,,
\label{eq2-10-2-2}
\end{align}
which describes the Dirac monopole \cite{Dirac:1931kp}. 
Note that $\bm{A}_{+}$ is well-defined when $\theta \neq \pi$, whereas $\bm{A}_{-}$ is well-defined when $\theta \neq 0$. In a common region, $\bm{A}_{+}$ and $\bm{A}_{-}$ are related through gauge transformation and hence they represent physically equivalent electromagnetic fields.

From the ansatz for the metric \eqref{eq2-4} and the configuration of the background magnetic field \eqref{eq2-5},
the Einstein equations \eqref{eq2-2} read
\begin{align}
{G_{t}}^{t} &= - \frac{1}{r^{2}} - \frac{h'}{rh^{2}} + \frac{1}{r^{2} h} = - 8 \pi G \mathcal{L} - \Lambda \,,
\label{eq2-10-3}\\
{G_{r}}^{r} &= - \frac{1}{r^{2}} + \frac{f'}{rfh} + \frac{1}{r^{2} h} = - 8 \pi G \mathcal{L} - \Lambda \,,
\label{eq2-10-4}\\
{G_{\theta}}^{\theta} =
{G_{\phi}}^{\phi} &= \frac{f''}{2fh} - \frac{{f'}^{2}}{4f^{2}h} - \frac{f'h'}{4fh^{2}} + \frac{f'}{2rfh} - \frac{h'}{2rh^{2}} 
= - 8 \pi G \mathcal{L} - \Lambda + \frac{8 \pi G q^{2} \mathcal{L}_{\mathcal{F}}}{r^{4}} 
\label{eq2-10-5} \, ,
\end{align}
where the right-hand sides are given by the background values, $\mathcal{L} = \mathcal{L}(q^{2}/(2r^{4}), 0)$ and $\mathcal{L}_{\mathcal{F}} = \mathcal{L}_{\mathcal{F}}(q^{2}/(2r^{4}), 0)$.
Subtracting Eq.\,\eqref{eq2-10-3} from Eq.\,\eqref{eq2-10-4} yields
\begin{align}
\frac{f'}{f} + \frac{h'}{h} = 0 \, ,
\label{eq2-11}
\end{align}
which implies $fh = \mathrm{constant}$.
Since we can always set the constant equal to one by rescaling the time coordinate, eventually we have
\begin{align}
h = \frac{1}{f} \, .
\label{eq2-12}
\end{align}
Then Eqs.\,\eqref{eq2-10-3} and \eqref{eq2-10-5} give
\begin{align}
- \frac{1}{r^{2}} + \frac{f'}{r} + \frac{f}{r^{2}} &= - 8 \pi G \mathcal{L} - \Lambda \, ,
\label{eq2-13}\\
\frac{f''}{2} + \frac{f'}{r}
&= - 8 \pi G \mathcal{L} - \Lambda + \frac{8 \pi G q^{2} \mathcal{L}_{\mathcal{F}}}{r^{4}}  \,.
\label{eq2-13-2}
\end{align}
We can get a solution to these equations as
\begin{align}
f(r) = 1 - \frac{2GM}{r} - \frac{\Lambda}{3} r^{2} - \frac{2GM_{q}(r)}{r} \, ,
\label{eq2-14}
\end{align}
where $M$ is an integration constant and we define
\begin{align}
M_{q}(r) \coloneqq 4 \pi \int dr \, r^{2} \mathcal{L}\left( \frac{q^{2}}{2r^{4}} , 0 \right)\,.
\label{eq2-15}
\end{align}
Notice that $M$ is the quantity corresponding to the mass of the black hole, and we are interested in the case $M > 0$.
Meanwhile, from the $t$ component of the equations of motion for the electromagnetic field \eqref{eq2-3}, we have
\begin{align}
\mathcal{L}_{\mathcal{F \widetilde{F}}}\left( \frac{q^{2}}{2r^{4}} , 0 \right) = 0 \, .
\label{eq2-16}
\end{align}
This relation constrains a form of $\mathcal{L}$
---when we expand $\mathcal{L}$ into power-series of $\widetilde{\mathcal{F}}$, the coefficient at the first-order must vanish:
\begin{align}
\mathcal{L}(\mathcal{F}, \widetilde{\mathcal{F}})
= \mathcal{L}_{0}(\mathcal{F})
+ \sum_{n=2}^{\infty} \frac{1}{n!} \mathcal{L}_{n}(\mathcal{F}) \widetilde{\mathcal{F}}^{n}\,,
\label{eq2-17}
\end{align}
where $\mathcal{L}_{0}$ and $\mathcal{L}_{n}$ are arbitrary functions of $\mathcal{F}$.

In the rest of this section, we check energy conditions which impose the reality and reasonability on the energy-momentum tensor.
Let us denote ${T_{t}}^{t} = - \rho$, ${T_{r}}^{r} = p_{r}$, ${T_{\theta}}^{\theta} = p_{\theta}$ and ${T_{\phi}}^{\phi} = p_{\phi}$,
where $\rho$ is the energy density and $p_{j} \, (j = r, \theta, \phi)$ correspond to the pressure along $j$-direction.
From Eq.\,\eqref{eq2-2-2}, these are given by
\begin{align}
\rho &= \mathcal{L} + \frac{\Lambda}{8 \pi G} \,,
\label{eq2-50}\\
p_{r} &= - \mathcal{L} - \frac{\Lambda}{8 \pi G} \,,
\label{eq2-51}\\
p_{\theta} = p_{\phi} &= - \mathcal{L} - \frac{\Lambda}{8 \pi G} + \frac{q^{2} \mathcal{L}_{\mathcal{F}}}{r^{4}}  \,,
\label{eq2-52}
\end{align}
where the right-hand sides are interpreted as the background values again, $\mathcal{L} = \mathcal{L}(q^{2}/(2r^{4}), 0)$ and $\mathcal{L}_{\mathcal{F}} = \mathcal{L}_{\mathcal{F}}(q^{2}/(2r^{4}), 0)$.
In terms of them, the statements of some energy conditions are as follows.
\begin{itemize}
\item
The weak energy condition --- which states $\rho \geq 0$ and $\rho + p_{j} \geq 0 ~ (j = r, \theta, \phi)$, for which we have
\begin{align}
\mathcal{L} + \frac{\Lambda}{8 \pi G} \geq 0
\quad \text{and} \quad 
\mathcal{L}_{\mathcal{F}} \geq 0 \,.
\label{eq2-53}
\end{align}

\item
The null energy condition --- $\rho + p_{j} \geq 0$,
\begin{align}
\mathcal{L}_{\mathcal{F}} \geq 0 \,.
\label{eq2-54}
\end{align}

\item
The dominant energy condition --- $\rho \geq |p_{j}|$,
\begin{align}
\mathcal{L} + \frac{\Lambda}{8 \pi G} \geq 0
 \,, \quad 
\mathcal{L}_{\mathcal{F}} \geq 0
\quad \text{and} \quad
2 \left( \mathcal{L} + \frac{\Lambda}{8 \pi G} \right) - \frac{q^{2} \mathcal{L}_{\mathcal{F}}}{r^{4}} \geq 0 \,.
\label{eq2-55}
\end{align}

\item
The strong energy condition --- $\rho + p_{j} \geq 0$ and $\rho + \sum_{j} p_{j} \geq 0$,
\begin{align}
\mathcal{L}_{\mathcal{F}} \geq 0
\quad \text{and} \quad
-2 \left( \mathcal{L} + \frac{\Lambda}{8 \pi G} \right) + \frac{q^{2} \mathcal{L}_{\mathcal{F}}}{r^{4}} \geq 0 \,.
\label{eq2-56}
\end{align}
\end{itemize}

\section{Master equations for black hole perturbations}
\label{sec:perturbations}

In this section, we derive the equations of motion for linear perturbations of metric and electromagnetic field on a background discussed in Section \ref{sec:background}.
Our goal is to obtain the master equations in the form of simultaneous Schr\"odinger-like equations with symmetric potentials.

In order to study the dynamics of metric and electromagnetic perturbations, we write
\begin{align}
g_{\mu\nu} &= \bar{g}_{\mu\nu} + \delta g_{\mu\nu} \, ,
\label{eq4-1}\\
F_{\mu\nu} &= \bar{F}_{\mu\nu} + \delta F_{\mu\nu} \, ,
\label{eq4-2}
\end{align}
where the quantities with a bar denote the spherically symmetric  background with a magnetic charge, which are studied in detail in Section \ref{sec:background}.
Here we recall them, 
\begin{align}
&\bar{g}_{\mu\nu}dx^{\mu}dx^{\nu} =  -f(r) \, dt^2 + \frac{1}{f(r)} \, dr^2 +  r^2 \, d\Omega^{2} \, ,
\label{eq4-3}\\
&\frac{1}{2}\bar{F}_{\mu\nu} dx^{\mu} \wedge dx^{\nu} = q \sin \theta \, d\theta \wedge d\phi \, ,
\label{eq4-4}
\end{align}
where $f(r)$ is a function of $r$ given by Eq.\,\eqref{eq2-14} and $q$ is a constant corresponding to a magnetic charge.
We take the background gauge potential as Eq.\,\eqref{eq2-10-2-2}, that is,
\begin{align}
\bar{A}_{\mu} dx^{\mu} = q (\pm 1 - \cos \theta) \, d\phi \,.
\label{eq4-4-2}
\end{align}
As we have seen in Eq.\,\eqref{eq2-10}, the invariants related to the background are given by
\begin{align}
\bar{\mathcal{F}} = \frac{q^{2}}{2r^{4}}\,, \qquad
\bar{\mathcal{\widetilde{F}}} = 0 \, .
\label{eq4-5}
\end{align}
On the other hand, $\delta g_{\mu\nu}$ and $\delta F_{\mu\nu}$ represent small fluctuations of metric and electromagnetic field strength, respectively.

By virtue of the spherical symmetry of the background, it is useful to expand the perturbations on a basis of tensor spherical harmonics to separate the radial and angular dependence of them.
After that, we can proceed to split the perturbations by their properties under parity transformation (spatial inversion),
\begin{align}
&\delta {g}_{\mu\nu} = \delta {g}_{\mu\nu}^{-} + \delta g^{+}_{\mu\nu} \, ,
\label{eq4-6}\\
&\delta F_{\mu\nu} = \delta {F}_{\mu\nu}^{-} + \delta F^{+}_{\mu\nu} \, ,
\label{eq4-6-2}
\end{align}
where $\delta {g}_{\mu\nu}^{-}$ and $ \delta {F}_{\mu\nu}^{-}$ represent the terms which pick a factor of $(-1)^{l+1}$ under parity transformation, where $l$ is the azimuthal quantum number of the spherical harmonics, and they are called to be \emph{odd}, \emph{axial}, or \emph{magnetic}.
On the other hand,  $\delta {g}_{\mu\nu}^{+}$ and $ \delta {F}_{\mu\nu}^{+}$ pick a factor of $(-1)^{l}$ under parity transformation, and are said to be \emph{even}, \emph{polar}, or \emph{electric}.
As we discuss in Appendix \ref{app:gauge}, choosing an appropriate gauge following Ref.\,\cite{Regge:1957td, Zerilli:1970se, Zerilli:1974ai}, we can write the metric perturbations as the following matrices with components in the $(t,r,\theta,\phi)$ coordinates,
\begin{align}
\delta g^{-}_{\mu\nu}&= \sum_{l,m}
\begin{pmatrix}
0	&0	&-[h_{0}/\sin \theta] \, \partial_{\phi}	&h_{0}\sin \theta \, \partial_{\theta}		\\
0	&0	&-[h_{1}/\sin \theta] \, \partial_{\phi}	&h_{1}\sin \theta \, \partial_{\theta}		\\
\star	&\star	&0	&0	\\
\star	&\star	&0	&0
\end{pmatrix}
Y_{lm} \, ,
\label{eq4-7}\\
\delta g^{+}_{\mu\nu}&= \sum_{l,m}
\begin{pmatrix}
fH_{0}	&H_{1}	&0	&0		\\
\star	&f^{-1}H_{2}	&0	&0		\\
0	&0	&r^{2}K	&0	\\
0	&0	&0	&r^{2}\sin^{2}\theta \, K
\end{pmatrix}
Y_{lm} \, ,
\label{eq4-8}
\end{align}
where $Y_{lm} = Y_{lm}(\theta, \phi)$ are  spherical harmonics, and $h_{0}$, $h_{1}$, $H_{0}$, $H_{1}$, $H_{2}$ and $K$ are functions of $(t,r)$ given for each mode $(l,m)$.
The functions $K$ and $h_{0}$ are defined for $l \geq 2$,
$H_{1}$ and $h_{1}$ are defined for $l \geq 1$, while $H_{0}$ and $H_{2}$ are defined for $l \geq 0$.
Note that the components denoted as stars can be obtained from the symmetry of the matrices.

Following Ref.\,\cite{Zerilli:1974ai}, we write down the perturbations of the electromagnetic field strength as
\begin{align}
\delta F^{-}_{\mu\nu}&= \sum_{l,m}
\begin{pmatrix}
0	&0	&[f^{-}_{02}/\sin \theta] \, \partial_{\phi}	&- f^{-}_{02} \sin \theta \, \partial_{\theta}		\\
0	&0	&[f^{-}_{12}/\sin \theta] \, \partial_{\phi}	&- f^{-}_{12} \sin \theta \, \partial_{\theta}		\\
\times	&\times&0	&f^{-}_{23} \sin \theta 	\\
\times	&\times&\times	&0
\end{pmatrix}
Y_{lm} \, ,
\label{eq4-9}\\
\delta F^{+}_{\mu\nu}&= \sum_{l,m}
\begin{pmatrix}
0	&f^{+}_{01}	&f^{+}_{02} \, \partial_{\theta}	&f^{+}_{02} \, \partial_{\phi}		\\
\times	&0	&f^{+}_{12} \, \partial_{\theta}	&f^{+}_{12} \, \partial_{\phi}		\\
\times&\times 	&0	&0	\\
\times &\times	&0	&0
\end{pmatrix}
Y_{lm} \, ,
\label{eq4-10}
\end{align}
where $f^{+}_{01}$, $f^{\pm}_{02}$, $f^{\pm}_{12}$ and $f^{-}_{23}$ are functions of $(t,r)$ given for each $(l,m)$.
The functions $f^{\pm}_{02}$, $f^{-}_{12}$ and $f^{-}_{23}$ are defined for $l \geq 1$, while $f^{+}_{01}$ and $f^{+}_{12}$ are defined for $l \geq 0$.
The cross marks stand for the components determined by the antisymmetry of the matrices.
We can regard $\delta F^{\pm}_{\mu\nu}$ as being derived from the perturbations of the vector potential $\delta A^{\pm}_{\mu}$ via
\begin{align}
\delta F^{\pm}_{\mu\nu} = \partial_{\mu} \delta A^{\pm}_{\nu} - \partial_{\nu} \delta A^{\pm}_{\mu} \,.
\label{eq4-11}
\end{align}
For the odd-parity part, the components of the potential can be expanded as
\begin{align}
&\delta A^{-}_{t} = 0 \,,
\label{eq4-12}\\
&\delta A^{-}_{r} = 0 \,,
\label{eq4-13}\\
&\delta A^{-}_{\theta} = \sum_{l,m} \frac{f^{-}_{23}}{l(l+1) \sin \theta}\partial_{\phi} Y_{lm} \,,
\label{eq4-14}\\
&\delta A^{-}_{\phi} = - \sum_{l,m} \frac{f^{-}_{23} }{l(l+1)} \sin \theta \,\partial_{\theta} Y_{lm} \, .
\label{eq4-15}
\end{align}
Then we can see the perturbation variables of the field strength are related by
\begin{align}
&f^{-}_{02} = \frac{\dot{f}^{-}_{23}}{l(l+1)} \, ,
\label{eq4-16}\\
&f^{-}_{12} = \frac{f^{-\prime}_{23}}{l(l+1)} \, ,
\label{eq4-17}
\end{align}
where a dot and a prime denote the derivative with respect to $t$ and $r$, respectively. 
For the even-parity part, the components of the potential are expanded as
\begin{align}
&\delta A^{+}_{t} = - \sum_{l,m} f^{+}_{02} Y_{lm} \,,
\label{eq4-18}\\
&\delta A^{+}_{r} = - \sum_{l,m} f^{+}_{12} Y_{lm} \,,
\label{eq4-19}\\
&\delta A^{+}_{\theta} = 0 \,,
\label{eq4-20}\\
&\delta A^{+}_{\phi} = 0 \,,
\label{eq4-21}
\end{align}
and the field strength perturbations are related by
\begin{align}
&f^{+}_{01} = f^{+\prime}_{02} - \dot{f}^{+}_{12} \, .
\label{eq4-22}
\end{align}

Notice that the background magnetic field is odd-parity as we can see from Eq.\,\eqref{eq4-4}, while the background metric \eqref{eq4-3} is even-parity.
(The spherically symmetric background corresponds to the mode $l=0$.)
Therefore, we can expect that the odd-parity parts of the metric perturbations couple to the even-parity parts of the electromagnetic perturbations to the linear order: we call this combination the type-I.
On the other hand,  the even-parity metric perturbations can couple to the odd-parity electromagnetic perturbations, which we call the type-II.

\subsection{Type-I: odd-parity metric and even-parity electromagnetic perturbations}

Here let us focus on the system type-I, in which the odd-parity metric and the even-parity electromagnetic perturbations are coupled.
First, we list the variables in the system below,
\begin{align}
\text{odd-parity} ~ \delta g^{-}_{\mu\nu}&:~ h_{0}\,, h_{1} \,,
\notag \\
\text{even-parity} ~ \delta F^{+}_{\mu\nu}&:~ f^{+}_{01} \,, f^{+}_{02} \,, f^{+}_{12} \,.
\notag
\end{align}

By substituting the expansions \eqref{eq4-1} and \eqref{eq4-2} with the odd-parity metric perturbations \eqref{eq4-7} and the even-parity electromagnetic perturbations \eqref{eq4-10} into the field equations \eqref{eq2-2} and \eqref{eq2-3}, and linearizing them, we can obtain the equations of motion for the type-I perturbations.
Here we write them as the Fourier transforms with respect to time $t$, that is, we replace $\partial_{t}$ by $- i \omega$.
The $({_{t}}^{\phi})$, $({_{r}}^{\phi})$ and $({_{\theta}}^{\phi})$ components of the Einstein equations give
\begin{align}
f h_{0}'' + i \omega f h_{1}'
- \left( f'' + \frac{l(l+1)}{r^{2}} \right) h_{0}
+ \frac{2 i \omega f}{r} h_{1}
&= \frac{16 \pi G q \mathcal{L}_{\mathcal{F}} }{r^{2}} f^{+}_{02}
\,,
\label{eq-Ein03}\\
%
\frac{i \omega}{f} h_{0}'
- \frac{2 i \omega}{rf} h_{0}
+ \left( f'' - \frac{\omega^{2}}{f} + \frac{l(l+1) - 2f}{r^{2}} \right) h_{1}
&= - \frac{16 \pi G q \mathcal{L}_{\mathcal{F}} }{r^{2}} f^{+}_{12}
\, ,
\label{eq-Ein13}\\
%
(f h_{1})' + \frac{i \omega {h}_{0}}{f}
&= 0
\,,
\label{eq-Ein23}
\end{align}
and from the $t$, $r$ and $\theta$ components of the gauge field equations, we have
\begin{align}
\left( \mathcal{L}_{\mathcal{F}} - \frac{q^{2}}{r^{4}} \mathcal{L}_{\mathcal{\widetilde{F}\widetilde{F}}} \right) f^{+\prime}_{01}
+ \frac{2}{r} \left[ \mathcal{L}_{\mathcal{F}} + \frac{q^{2}}{r^{4}} \left( \mathcal{L}_{\mathcal{\widetilde{F}\widetilde{F}}} - \mathcal{L}_{\mathcal{FF}} \right) + \frac{q^{4}}{r^{8}} \mathcal{L}_{\mathcal{F\widetilde{F}\widetilde{F}}} \right] f^{+}_{01}
- \frac{l(l+1) \mathcal{L}_{\mathcal{F}}}{r^{2}f} f^{+}_{02}
- \frac{l(l+1) q \mathcal{L}_{\mathcal{F}}}{r^{4}f} h_{0}
&= 0 
\,,
\label{eq-Max0}\\
i \omega \left( \mathcal{L}_{\mathcal{F}} - \frac{q^{2}}{r^{4}} \mathcal{L}_{\mathcal{\widetilde{F}\widetilde{F}}} \right) f^{+}_{01}
+ \frac{f l(l+1) \mathcal{L}_{\mathcal{F}}}{r^{2}} f^{+}_{12}
+ \frac{f l(l+1) q \mathcal{L}_{\mathcal{F}}}{r^{4}} h_{1}
&= 0
\,,
\label{eq-Max1}\\
\mathcal{L}_{\mathcal{F}} f^{+\prime}_{12}
+ \left( \frac{f' \mathcal{L}_{\mathcal{F}}}{f} - \frac{2q^{2} \mathcal{L}_{\mathcal{FF}}}{r^{5}} \right) f^{+}_{12}
+ \frac{i \omega \mathcal{L}_{\mathcal{F}}}{f^{2}} f^{+}_{02}
+ \frac{q \mathcal{L}_{\mathcal{F}}}{r^{2}} h_{1}'
+ \left[
\frac{q \mathcal{L}_{\mathcal{F}}}{r^{2}} \left( \frac{f'}{f} - \frac{2}{r} \right) - \frac{2q^{3} \mathcal{L}_{\mathcal{FF}}}{r^{7}}
\right] h_{1}
+ \frac{i \omega q \mathcal{L}_{\mathcal{F}}}{r^{2} f^{2}} h_{0}
&= 0 \,,
\label{eq-Max2}
\end{align}
where $\mathcal{L}_{\mathcal{F}}$ means the background value $\mathcal{L}_{\mathcal{F}}(\bar{\mathcal{F}}, \bar{\mathcal{\widetilde{F}}})$, and so on.
The equations \eqref{eq-Max0} and \eqref{eq-Max1} hold for $l \geq 0$,
Eqs.\,\eqref{eq-Ein03}, \eqref{eq-Ein13} and \eqref{eq-Max2} hold for $l \geq 1$, while Eq.\,\eqref{eq-Ein23} holds for $l \geq 2$.
In fact, Eq.\,\eqref{eq-Ein03} is redundant since it is automatically satisfied by using Eqs.\,\eqref{eq-Ein13}, \eqref{eq-Ein23} and \eqref{eq-Max2} with the background equations \eqref{eq2-13} and \eqref{eq2-13-2}.
The equation \eqref{eq-Max2} is also redundant because we can combine Eq.\,\eqref{eq-Max0} with Eq.\,\eqref{eq-Max1} to obtain it.
Therefore, we below utilize four equations \eqref{eq-Ein13}--\eqref{eq-Max1} and an identity \eqref{eq4-22} for five variables $h_{0}$, $h_{1}$, $f^{+}_{01}$, $f^{+}_{02}$ and $f^{+}_{12}$ to get a couple of ``master equations'' which is a system of Schr\"odinger-like second-order differential equations and determines the dynamics of the perturbations.

For mode $l = 0$ corresponding to the spherically symmetric case, the equations above do not describe the degree of freedom propagating spacetime, thus we are not interested in the mode.
We first focus on the modes $l \geq 2$, in which metric and electromagnetic perturbations are coupled together.
After that, we turn to the modes $l = 1$, in which only the electromagnetic perturbations exist as the physical degrees of freedom.

\subsubsection{Type-I master equations for $l \geq 2$}

We can solve Eq.\,\eqref{eq-Ein23} for $h_{0}$ and Eq.\,\eqref{eq-Max1} for $f^{+}_{12}$, and substitute them into Eq.\,\eqref{eq-Ein13}.
Then we get a second-order differential equation for $h_{1}$ in terms of $f^{+}_{01}$.
On the other hand, we can obtain a second-order equation for $f^{+}_{01}$ in terms of $h_{1}$ by solving Eq.\,\eqref{eq-Max0} for $f^{+}_{02}$ and Eq.\,\eqref{eq-Max1} for $f^{+}_{12}$ and substituting them into Eq.\,\eqref{eq4-22}.
Let us define ``master variables'' by
\begin{align}
\mathcal{R}^{-} 
&\coloneqq \frac{fh_{1}}{ \sqrt{8 \pi G} (-i \omega) r} \, ,
\label{eq-type1-1}\\
\mathcal{E} 
&\coloneqq \frac{ \sqrt{2} \, r^{2} }{l(l+1) \sqrt{(l+2)(l-1) |\mathcal{L}_{\mathcal{F}}| } } 
\left( \mathcal{L}_{\mathcal{F}} - \frac{q^{2}}{r^{4}} \mathcal{L}_{\mathcal{\widetilde{F}\widetilde{F}}} \right)
f^{+}_{01}
\, ,
\label{eq-type1-2}
\end{align}
and use the tortoise coordinate $r^{*}$ which is defined by $f dr^{*} = dr$.
Then we have a couple of master equations,
\begin{align}
\frac{d^{2}\mathcal{R}^{-}}{dr^{*2}} 
+ \left[ \omega^{2} 
- f \left\{
8 \pi G \mathcal{L} + \Lambda
+ \frac{l(l+1) + 3(f-1)}{r^{2}}
\right\}
\right] \mathcal{R}^{-}
- \frac {fq \sqrt {16 \pi G (l+2)(l-1) | \mathcal{L}_{\mathcal{F}} | }}{{r}^{3}} \mathcal{E}&=0 \,,
\label{eq-type1-3}\\
\frac{d^{2}\mathcal{E}}{dr^{*2}} 
+ \bigg[ \omega^{2}
- f \bigg\{
{\frac{l(l+1)}{r^{2}} \frac{ \mathcal{L}_{\mathcal{F}}}{\mathcal{L}_{\mathcal{F}}- (q^{2}/r^{4}) \mathcal{L}_{\mathcal{\widetilde{F}\widetilde{F}}}}} 
+{\frac {16\pi G q^{2} \mathcal{L}_{\mathcal{F}}}{{r}^{4}}}
-\left( 8 \pi G \mathcal{L} + \Lambda + \frac{6f -1}{r^{2}} \right) \frac{q^{2} \mathcal{L}_{\mathcal{FF}}}{r^{4} \mathcal{L}_{\mathcal{F}}}
&\notag \\
+{\frac { 3 f q^{4} \mathcal{L}_{\mathcal{FF}}^{2} }{ r^{10} \mathcal{L}_{\mathcal{F}}^{2} }}
-{\frac {2 f q^{4} \mathcal{L}_{\mathcal{FFF}} }{r^{10} \mathcal{L}_{\mathcal{F}} }} 
\bigg\} \bigg] \mathcal{E} 
- \mathrm{sgn}(\mathcal{L}_{\mathcal{F}})  \frac {fq  \sqrt { 16 \pi G (l+2)(l-1) | \mathcal{L}_{\mathcal{F}} | }}{{r}^{3}} \mathcal{R}^{-} &= 0 \,,
\label{eq-type1-4}
\end{align}
where we utilized the background equations \eqref{eq2-13} and \eqref{eq2-13-2}.
These equations can be rewritten as
\begin{align}
\frac{1}{f} \left[ 
r \frac{d}{dr^{*}} \left( \frac{1}{r^{2}} \frac{d}{dr^{*}} \left( r \mathcal{R}^{-} \right) \right)
+ \omega^{2} \mathcal{R}^{-}
\right]
- V^{\mathrm{I}}_{11} \mathcal{R}^{-}
- V^{\mathrm{I}}_{12} \mathcal{E} &=0 \, ,
\label{eq-type1-5}\\
\frac{1}{f} \left[ 
\sqrt{|\mathcal{L}_{\mathcal{F}}|} \frac{d}{dr^{*}} \left( \frac{1}{\mathcal{L}_{\mathcal{F}}} \frac{d}{dr^{*}} \left( \sqrt{|\mathcal{L}_{\mathcal{F}}|} \mathcal{E} \right) \right)
+ \mathrm{sgn}(\mathcal{L}_{\mathcal{F}}) \, \omega^{2} \mathcal{E}
\right]
- V^{\mathrm{I}}_{22} \mathcal{E}
- V^{\mathrm{I}}_{21} \mathcal{R}^{-} &=0 
\,,
\label{eq-type1-6}
\end{align}
where
\begin{align}
V^{\mathrm{I}}_{11} &= \frac{(l+2)(l-1)}{r^{2}} \,,
\label{eq-type1-7}\\
V^{\mathrm{I}}_{12} = V^{\mathrm{I}}_{21} &= \frac {q \sqrt {16 \pi G (l+2)(l-1) | \mathcal{L}_{\mathcal{F}}| }}{{r}^{3}} \,,
\label{eq-type1-8}\\
V^{\mathrm{I}}_{22} &=
|\mathcal{L}_{\mathcal{F}}|
\left[
\frac{l(l+1)}{r^{2}} \frac{1}{\mathcal{L}_{\mathcal{F}} - (q^{2}/r^{4}) \mathcal{L}_{\mathcal{\widetilde{F}\widetilde{F}}} } 
+ \frac{16\pi G q^{2} }{r^{4}} 
\right]
\,.
\label{eq-type1-9}
\end{align}
Getting back to the time domain, the quadratic action for the perturbations which induces the equations of motion above is given by
\begin{align}
S_{\mathrm{I}}[\mathcal{R}^{-}, \mathcal{E}] 
= \int dt dr^{*} f \, \bigg[
&- \frac{1}{r^{2}} 
 \nabla^{a} (r \mathcal{R}^{-})^{*}
 \nabla_{a} (r \mathcal{R}^{-})
- \frac{1}{\mathcal{L}_{\mathcal{F}}}
 \nabla^{a} \left( \sqrt{|\mathcal{L}_{\mathcal{F}}|} \mathcal{E} \right)^{*}
 \nabla_{a} \left( \sqrt{|\mathcal{L}_{\mathcal{F}}|} \mathcal{E} \right)
- 
\begin{pmatrix}
\mathcal{R}^{-*} & \mathcal{E}^{*} 
\end{pmatrix} 
\bm{V}_{\mathrm{I}}
\begin{pmatrix}
\mathcal{R}^{-} \\
\mathcal{E} 
\end{pmatrix} 
\bigg] \,,
\label{eq4-1-6-3}
\end{align}
where an index $a$ runs over $(t, r^{*})$,
and we introduce the potential matrix as
\begin{align}
\bm{V}_{\mathrm{I}} =
\begin{pmatrix}
V^{\mathrm{I}}_{11}	&V^{\mathrm{I}}_{12} \\
\star	&V^{\mathrm{I}}_{22}
\end{pmatrix} \,.
\label{eq4-1-7}
\end{align} 
Note that the volume element in Eq.\,\eqref{eq4-1-6-3} satisfies $dt dr^{*} f = dt dr $.
We can find that the necessary condition to avoid the existence of a ghost is
\begin{align}
\mathcal{L}_{\mathcal{F}} > 0 \,.
\label{eq4-1-11} 
\end{align}
Note that this condition is consistent with the energy conditions considered in Eqs.\,\eqref{eq2-53}--\eqref{eq2-56}.

Given the solutions for the master equations, $\mathcal{R}^{-}$ and $\mathcal{E}$, the metric and electromagnetic perturbations in Fourier space can be obtained by
\begin{align}
h_{1} &= - \frac{i \omega \sqrt{8 \pi G} }{f} r \mathcal{R}^{-} \, ,
\label{eq4-1-20}\\
h_{0} &= \sqrt{8 \pi G} f (r \mathcal{R}^{-})' \,,
\label{eq4-1-21}\\
f^{+}_{01} &= \frac{1}{\sqrt{2} r^{2}} \frac{ l (l+1) \sqrt{(l+2)(l-1) } }{ \mathcal{L}_{\mathcal{F}} - (q^{2}/r^{4}) \mathcal{L}_{\mathcal{\widetilde{F}\widetilde{F}}} } \sqrt{\mathcal{L}_{\mathcal{F}}} \mathcal{E} \,,
\label{eq4-1-22}\\
f^{+}_{02} &= \frac{f \sqrt{(l+2) (l-1)}}{\sqrt{2} \mathcal{L}_{\mathcal{F}}} \left( \sqrt{\mathcal{L}_{\mathcal{F}}} \mathcal{E} \right)'
- \frac{ \sqrt{8 \pi G} fq}{r^{2}} (r \mathcal{R}^{-})' \,,
\label{eq4-1-23}\\
f^{+}_{12} &= - \frac{i \omega \sqrt{(l+2)(l-1) }}{\sqrt{2} f \mathcal{L}_{\mathcal{F}} } \sqrt{\mathcal{L}_{\mathcal{F}}} \mathcal{E}
+ \frac{i \omega \sqrt{8 \pi G} q}{r f} \mathcal{R}^{-} \,.
\label{eq4-1-24}
\end{align}

\subsubsection{Type-I master equations for $l=1$}

For the modes $l = 1$, the variable $h_{0}$ is not defined, and only the electromagnetic perturbations are dynamical degrees of freedom.
We define the master variable for $l = 1$ as
\begin{align}
\mathcal{E} \coloneqq \frac{r^{2}}{\sqrt{\mathcal{L}_{\mathcal{F}}}} \left( \mathcal{L}_{\mathcal{F}} - \frac{q^{2}}{r^{4}} \mathcal{L}_{\mathcal{\widetilde{F}\widetilde{F}}} \right) f^{+}_{01} \,,
\label{eq4-1-2-1}
\end{align}
where we think of $\mathcal{L}_{\mathcal{F}}$ as positive since that is necessary for ghost-freeness as seen in the modes $l \geq 2$.
Then we have the following master equation,
\begin{align}
\frac{d^{2}\mathcal{E}}{dr^{*2}} 
+ \bigg[ \omega^{2}
- f \bigg\{
\frac{2}{r^{2}} \frac{\mathcal{L}_{\mathcal{F}}}{\mathcal{L}_{\mathcal{F}} - (q^{2}/r^{4}) \mathcal{L}_{\mathcal{\widetilde{F}\widetilde{F}}} }
+ \frac{16 \pi G q^{2} \mathcal{L}_{\mathcal{F}}}{r^{4}}
- \left( 8 \pi G \mathcal{L} + \Lambda + \frac{6f - 1}{r^{2}} \right)
\frac{q^{2} \mathcal{L}_{\mathcal{FF}}}{r^{4} \mathcal{L}_{\mathcal{F}}}
\qquad&\notag \\
+ \frac{3fq^{4} \mathcal{L}_{\mathcal{FF}}^{2}}{r^{10} \mathcal{L}_{\mathcal{F}}^{2}}
- \frac{2 f q^{4} \mathcal{L}_{\mathcal{FFF}}}{r^{10} \mathcal{L}_{\mathcal{F}}}
\bigg\} \bigg] \mathcal{E} 
&= 0 \,,
\label{eq4-1-2-2}
\end{align}
which can be rewritten as 
\begin{align}
\frac{1}{f} \left[ \sqrt{\mathcal{L}_{\mathcal{F}}} \frac{d}{dr^{*}} \left( \frac{1}{\mathcal{L}_{\mathcal{F}}} \frac{d}{dr^{*}} \left( \sqrt{\mathcal{L}_{\mathcal{F}}} \mathcal{E} \right) \right) 
+ \omega^{2} \mathcal{E}
\right]
- \left( \frac{2}{r^{2}} \frac{\mathcal{L}_{\mathcal{F}}}{\mathcal{L}_{\mathcal{F}} - (q^{2}/r^{4}) \mathcal{L}_{\mathcal{\widetilde{F}\widetilde{F}}}}
+ \frac{16 \pi G q^{2} \mathcal{L}_{\mathcal{F}}}{r^{4}} \right) \mathcal{E} = 0 \,.
\label{eq4-1-2-3}
\end{align}
In terms of $\mathcal{E}$, the electromagnetic field perturbations are given by
\begin{align}
f^{+}_{01} &= \frac{1}{r^{2}} \frac{1}{ \mathcal{L}_{\mathcal{F}} - (q^{2}/r^{4}) \mathcal{L}_{\mathcal{\widetilde{F}\widetilde{F}}} } \sqrt{\mathcal{L}_{\mathcal{F}}} \mathcal{E} \,,
\label{eq4-1-2-4}\\
f^{+}_{02} &= \frac{f}{2 \mathcal{L}_{\mathcal{F}}} 
\left( \sqrt{\mathcal{L}_{\mathcal{F}}} \mathcal{E} \right)'
\,,
\label{eq4-1-2-5}\\
f^{+}_{12} &= - \left( \frac{i \omega}{2f \mathcal{L}_{\mathcal{F}}} + \frac{8 \pi G q^{2}}{i \omega r^{4}} \right) \sqrt{\mathcal{L}_{\mathcal{F}}} \mathcal{E}
\,.
\label{eq4-1-2-6}
\end{align}
Now the metric perturbation $h_{1}$ is not a physical degree of freedom and determined by the electromagnetic perturbation. 
It is given by
\begin{align}
h_{1} &= \frac{8 \pi G q}{i \omega r^{2}} \sqrt{\mathcal{L}_{\mathcal{F}}} \mathcal{E} \,.
\label{eq-4-1-2-7}
\end{align}

\subsection{Type-II: even-parity metric and odd-parity electromagnetic perturbations}

Let us turn to the system type-II, in which the even-parity metric and the odd-parity electromagnetic perturbations form a coupled system together.
We deal with the following variables:
\begin{align}
\text{even-parity} ~ \delta g^{+}_{\mu\nu}&:~ H_{0}\,, H_{1} \,, H_{2} \,, K \,,
\notag \\
\text{odd-parity} ~ \delta F^{-}_{\mu\nu}&:~ f^{-}_{02} \,, f^{-}_{12} \,, f^{-}_{23} \,.
\notag
\end{align} 

From the $({_{t}}^{t})$, $({_{t}}^{r})$, $({_{r}}^{r})$, $({_{t}}^{\phi})$, $({_{r}}^{\phi})$, $({_{\theta}}^{\theta})$ and $({_{\theta}}^{\phi})$ components of the Einstein equations, we have
\begin{align}
f K'' + \left( \frac{f'}{2} + \frac{3f}{r} \right) K'
- \frac{(l+2)(l-1)}{2r^{2}} K 
- \frac{f}{r} H_{2}'
- \frac{l(l+1)}{2r^{2}} H_{2} - \frac{(rf)'}{r^{2}} H_{2}
&= \frac{8 \pi G q^{2} \mathcal{L}_{\mathcal{F}} }{r^{4}}
\left( K - \frac{f^{-}_{23}}{q} \right) 
\, ,
\label{eq-2-Ein00}\\
%
i \omega f K' + \frac{i \omega (2f - rf')}{2r} K
- \frac{i \omega f }{r} H_{2} + \frac{fl(l+1)}{2r^{2}} H_{1}
&= 0
\, ,
\label{eq-2-Ein01}\\
%
\left( \frac{f'}{2} + \frac{f}{r} \right) K'
- \frac{(l+2)(l-1)}{2r^{2}} K + \frac{\omega^{2}}{f} K 
- \frac{f}{r} H_{0}'
+ \frac{l(l+1)}{2r^{2}} H_{0} - \frac{2 i \omega}{r} H_{1}
- \frac{(rf)'}{r^{2}} H_{2}
&= \frac{8 \pi G q^{2} \mathcal{L}_{\mathcal{F}} }{r^{4}} \left( K - \frac{f^{-}_{23}}{q} \right) 
\, ,
\label{eq-2-Ein11}\\
%
f H_{1}' + f' H_{1} + i \omega K + i \omega H_{2}
&= - \frac{16 \pi G q \mathcal{L}_{\mathcal{F}} }{r^{2}} f^{-}_{02} 
\, ,
\label{eq-2-Ein03}\\
%
K' - H_{0}' + \left( \frac{1}{r} - \frac{f'}{2f} \right) H_{0}
- \frac{i \omega}{f} H_{1} - \left( \frac{1}{r} + \frac{f'}{2 f} \right) H_{2}
&= \frac{16 \pi G q \mathcal{L}_{\mathcal{F}} }{r^{2}} f^{-}_{12}
\, ,
\label{eq-2-Ein13}\\
%
fK'' + \left( f' + \frac{2f}{r} \right) K' + \frac{\omega^{2}}{f} K
- f H_{0}'' - \left( \frac{3f'}{2} + \frac{f}{r} \right) H_{0}'
+ \frac{l(l+1)}{2r^{2}} H_{0} - 2 i \omega H_{1}' 
&\notag \\
- i \omega \left( \frac{2}{r} + \frac{f'}{f} \right) H_{1}
- \left( \frac{f'}{2} + \frac{f}{r} \right) H_{2}'
+ \left( \frac{\omega^{2}}{f} - f'' - \frac{2f'}{r} - \frac{l(l+1)}{2r^{2}} \right) H_{2}
&= \frac{16 \pi G q^{2}}{r^{4}} \left( \mathcal{L}_{\mathcal{F}} + \frac{q^{2}}{r^{4}} \mathcal{L}_{\mathcal{FF}} \right)
\notag \\
&\quad 
\times \left( \frac{f^{-}_{23}}{q} - K \right)
\, ,
\label{eq-2-Ein22}\\
%
H_{0} - H_{2}
&= 0
\, ,
\label{eq-2-Ein23}
\end{align}
where Eqs.\,\eqref{eq-2-Ein00}--\eqref{eq-2-Ein11}, \eqref{eq-2-Ein22} hold for $l \geq 0$, Eqs.\,\eqref{eq-2-Ein03} and \eqref{eq-2-Ein13} hold for $l \geq 	1$, and Eq.\,\eqref{eq-2-Ein23} holds for $l \geq 2$.
On the other hand, from the $\theta$ component of the electromagnetic field equations, we have
\begin{align}
f \mathcal{L}_{\mathcal{F}} f^{-\prime}_{12}
+ \left( f'\mathcal{L}_{\mathcal{F}} - \frac{2fq^{2} \mathcal{L}_{\mathcal{FF}}}{r^{5}} \right) f^{-}_{12}
+ \frac{i \omega \mathcal{L}_{\mathcal{F}}}{f} f^{-}_{02}
- \frac{1}{r^{2}} \left( \mathcal{L}_{\mathcal{F}} + \frac{q^{2}}{r^{4}} \mathcal{L}_{\mathcal{FF}} \right) f^{-}_{23}
&\notag \\
+ \frac{q}{r^{2}} \left( \mathcal{L}_{\mathcal{F}} + \frac{q^{2}}{r^{4}} \mathcal{L}_{\mathcal{FF}} \right) K
+ \frac{q \mathcal{L}_{\mathcal{F}}}{2r^{2}} (H_{0} - H_{2}) &= 0 \,,
\label{eq-2-Max2}
\end{align}
which holds for $l \geq 1$.
In the same manner as for the type-I, we first focus on the modes $l \geq 2$, then we turn to the modes $l = 1$.

\subsubsection{Type-II master equations for $l \geq 2$}

The equation \eqref{eq-2-Ein23} immediately implies that we can replace $H_{0}$ appearing in all other equations by $H_{2}$.
We can also use Eqs.\,\eqref{eq4-16} and \eqref{eq4-17} to express $f^{-}_{02}$ and $f^{-}_{12}$ in terms of $f^{-}_{23}$.
Then Eq.\,\eqref{eq-2-Max2} gives a second-order differential equation with respect to $f^{-}_{23}$,
\begin{align}
f^{-\prime\prime}_{23} 
+ \left( \frac{f'}{f} - \frac{2q^{2} \mathcal{L}_{\mathcal{FF}}}{r^{5} \mathcal{L}_{\mathcal{F}}} \right) f^{-\prime}_{23}
+ \left[
\frac{\omega^{2}}{f^{2}} - \frac{l(l+1)}{r^{2} f} - \frac{l(l+1) q^{2} \mathcal{L}_{\mathcal{FF}}}{r^{6} f \mathcal{L}_{\mathcal{F}}}
\right] f^{-}_{23}
+ \left[
\frac{l(l+1) q}{r^{2}f} + \frac{l(l+1) q^{3} \mathcal{L}_{\mathcal{FF}}}{r^{6} f \mathcal{L}_{\mathcal{F}}}
\right] K
= 0 \,,
\label{eq4-2-0}
\end{align}
which describes the dynamics of the electromagnetic perturbations.
Next, let us turn to deformations of the linearized Einstein equations.
We can solve Eq.\,\eqref{eq-2-Ein01} for $K'$ and substitute it into Eq.\,\eqref{eq-2-Ein13}.
Furthermore, Eqs.\,\eqref{eq-2-Ein23} and \eqref{eq4-17} allow us to replace $H_{0}$ by $H_{2}$ and $f^{-}_{12}$ by $f^{-\prime}_{23} / l(l+1)$.
Then Eq.\,\eqref{eq-2-Ein13} reduces to a first-order equation for $H_{2}$,
\begin{align}
H_{2}' + \left( \frac{f'}{f} - \frac{1}{r} \right) H_{2}
+ \left[ \frac{i \omega}{f} + \frac{l(l+1)}{2i \omega r^{2}} \right] H_{1}
+ \left( \frac{1}{r} - \frac{f'}{2f} \right) K
= - \frac{16 \pi G q \mathcal{L}_{\mathcal{F}}}{r^{2} l(l+1)} f^{-\prime}_{23} \,.
\label{eq4-2-1}
\end{align}
On the other hand, Eq.\,\eqref{eq4-16} can be used to rewrite Eq.\,\eqref{eq-2-Ein03} as
\begin{align}
f H_{1}' + f' H_{1} + i \omega K + i \omega H_{2}
&= \frac{16 \pi G i \omega q \mathcal{L}_{\mathcal{F}} }{r^{2} l(l+1)} f^{-}_{23} 
\, .
\label{eq4-2-1-2}
\end{align}
Thus we have three first-order differential equations \eqref{eq4-2-1-2}, \eqref{eq4-2-1} and \eqref{eq-2-Ein01} for three metric perturbations $H_{1}$, $H_{2}$ and $K$, to which the electromagnetic perturbation $f^{-}_{23}$ contributes as a source.
As seen below, we can further reduce the variables.
Let us substitute $K'$ from Eq.\,\eqref{eq-2-Ein01} into Eq.\,\eqref{eq-2-Ein11} and replace $H_{0}$ by $H_{2}$ from Eq.\,\eqref{eq-2-Ein23}.
Now we can eliminate $H_{2}'$ by using Eq.\,\eqref{eq4-2-1}. Then we find that Eq.\,\eqref{eq-2-Ein11} gives an algebraic relation between some variables,
\begin{align}
- \left[ \frac{2i\omega}{r} + \frac{f' l(l+1)}{2 i \omega r^{2}} \right] H_{1}
+ \frac{rf' - 2f + l(l+1)}{r^{2}} H_{2}
+ \left[
\frac{f^{\prime 2} + 4 \omega^{2}}{2f} - \frac{f'}{r} - \frac{(l+2)(l-1)}{r^{2}} - \frac{16 \pi G q^{2} \mathcal{L}_{\mathcal{F}}}{r^{4}}
\right] K
&
\notag \\
+ \frac{32 \pi G f q \mathcal{L}_{\mathcal{F}}}{r^{3} l(l+1)} f^{-\prime}_{23}
+ \frac{16 \pi G q \mathcal{L}_{\mathcal{F}}}{r^{4}} f^{-}_{23}
&= 0 \,.
\label{eq4-2-2}
\end{align}
We can use this relation \eqref{eq4-2-2} to remove one of the variables $H_{1}$, $H_{2}$ or $K$ from Eqs.\,\eqref{eq4-2-1-2}, \eqref{eq4-2-1} and \eqref{eq-2-Ein01}.
Now let us eliminate $H_{2}$, then Eqs.\,\eqref{eq-2-Ein01} and \eqref{eq4-2-1-2} reduce to
\begin{align}
K' + a_{1}(r) K + a_{2}(r) H_{1} &= j_{1} \,,
\label{eq4-2-3}\\
H_{1}' + b_{1}(r) K + b_{2}(r) H_{1} &= j_{2} \,,
\label{eq4-2-4}
\end{align}
where
\begin{align}
a_{1}(r) &= 
\frac{- 16 \pi G f q^{2} \mathcal{L}_{\mathcal{F}}
+ r^{2} \zeta(r) (f - \lambda - 1)
+ 2 r^{2} (\lambda + 1)^{2}
- 2 r^{2} f (2 \lambda + 1)
+ 2 \omega^{2} r^{4}
}{r^{3} f \zeta(r)} \,,
\label{eq4-2-5}\\
a_{2}(r) &= 
2 \frac{-f(\lambda + 1) + (\lambda + 1)^{2} + \omega^{2} r^{2}}{i \omega r^{2} \zeta(r)}\,,
\label{eq4-2-6}\\
b_{1}(r) &=
 i \omega 
\frac{
32 \pi G f q^{2} \mathcal{L}_{\mathcal{F}} 
- r^{2} \zeta^{2}(r) + 4 r^{2} \zeta(r) (\lambda +1)
+ 4 r^{2} f (2 \lambda +1 )
- 4 r^{2} (\lambda +1)^{2}
- 4 \omega^{2} r^{4}
}{2r^{2} f^{2} \zeta(r)} \,,
\label{eq4-2-7}\\
b_{2}(r) &= 
\frac{\zeta^{2}(r) - \zeta(r) (\lambda -2 f +1)  + 2(\lambda +1)(f - \lambda - 1) - 2 \omega^{2} r^{2}}{r f \zeta(r)} \,,
\label{eq4-2-8}
\end{align}
and
\begin{align}
j_{1} &= 
- \frac{16 \pi G q \mathcal{L}_{\mathcal{F}} [rf f^{-\prime}_{23} + (\lambda + 1) f^{-}_{23}] }{r^{3} \zeta(r) (\lambda + 1)} \,,
\label{eq4-2-9}  \\
j_{2} &= 
i \omega \frac{16 \pi G q \mathcal{L}_{\mathcal{F}}
[2 r f f^{- \prime}_{23} + (\zeta(r) + 2 \lambda + 2) f^{-}_{23}]
}{2 r^{2} f \zeta(r) (\lambda +1)} \,,
\label{eq4-2-10}
\end{align}
where we defined
\begin{align}
\lambda &\coloneqq \frac{1}{2}(l+2)(l-1) \,,
\label{eq4-2-16}\\
\zeta (r) &\coloneqq 
r f' -2 f + l(l+1) \notag \\
&= - r^{2} ( 8 \pi G \mathcal{L} + \Lambda ) - 3f + 2\lambda  +3 \,.
\label{eq4-2-17}
\end{align}
In the second line of Eq.\,\eqref{eq4-2-17}, we used the background equation \eqref{eq2-13}.
We have not used Eqs.\,\eqref{eq-2-Ein00} and \eqref{eq-2-Ein22} so far.
In fact, they are redundant equations since they are automatically satisfied by using other equations.

We can combine the couple of first-order equations \eqref{eq4-2-3} and 	\eqref{eq4-2-4} to get a single second-order equation.
We find that the following choice of master variables makes the resulting equation simple:
\begin{align}
\mathcal{R}^{+}
&\coloneqq
\frac{2r (\lambda + 1) \sqrt{2 \lambda}}{\zeta(r)} K
+ \frac{2f (\lambda +1) \sqrt{2 \lambda}}{i \omega \zeta(r)} H_{1} \,,
\label{eq4-2-11}\\
\mathcal{B}
&\coloneqq 
\sqrt{16 \pi G |\mathcal{L}_{\mathcal{F}}|}
\left(
f^{-}_{23} - \frac{2q (\lambda +1)}{\zeta(r)} K - \frac{2fq (\lambda +1)}{i \omega r \zeta(r)} H_{1}
\right) \,.
\label{eq4-2-12}
\end{align}
Using the tortoise coordinate $r^{*}$, we have a pair of the master equations for the type-II perturbations as
\begin{align}
\frac{d^{2} \mathcal{R}^{+}}{dr^{*2}} 
+ \left[ \omega^{2} 
- f \left\{
\frac{64 \pi G f q^{2} \lambda \mathcal{L}_{\mathcal{F}}}{r^{4} \zeta^{2}(r)}
+ \frac{\zeta(r)}{r^{2}} 
- \frac{2(2\lambda - f + 1)}{r^{2}}
+ \frac{8 \lambda (\lambda - f + 1)}{r^{2} \zeta(r)}
+ \frac{8 \lambda^{2} f}{r^{2} \zeta^{2}(r)}
\right\}
\right]\mathcal{R}^{+} 
&\notag \\
- \mathrm{sgn}(\mathcal{L}_{\mathcal{F}}) \sqrt{32 \pi G \lambda |\mathcal{L}_{\mathcal{F}}|} f q
\left[ 
- \frac{32 \pi G f q^{2} \mathcal{L}_{\mathcal{F}}}{r^{5} \zeta^{2}(r)}
- \frac{2 f q^{2} \mathcal{L}_{\mathcal{FF}}}{r^{7} \zeta(r) \mathcal{L}_{\mathcal{F}}}
+ \frac{1}{r^{3}} \left( 1 - \frac{2(2\lambda - f + 2)}{\zeta(r)} - \frac{4 \lambda f}{\zeta^{2}(r)} \right)
\right]
\mathcal{B} &= 0
\,,
\label{eq4-2-14}\\
\frac{d^{2} \mathcal{B}}{dr^{*2}} +
\left[ \omega^{2}
- f \left\{
\frac{8 (8 \pi G)^{2} fq^{4} \mathcal{L}_{\mathcal{F}}^{2} }{r^{6} \zeta^{2}(r)}
+ \frac{64 \pi G f q^{4} \mathcal{L}_{\mathcal{FF}}}{r^{8} \zeta(r)}
- \frac{16 \pi G q^{2} \mathcal{L}_{\mathcal{F}}}{r^{4}}
\left( 1 - \frac{4(\lambda+1)}{\zeta(r)} - \frac{4 \lambda f}{\zeta^{2}(r)} \right)
\right. \right. 
&\notag \\
\left. \left.
- \frac{q^{2} \mathcal{L}_{\mathcal{FF}}}{r^{6} \mathcal{L}_{\mathcal{F}}} \left( \zeta(r) - 3f - 4(\lambda + 1) \right)
+ \frac{2 f q^{4} \mathcal{L}_{\mathcal{FFF}}}{r^{10} \mathcal{L}_{\mathcal{F}}}
- \frac{f q^{4} \mathcal{L}_{\mathcal{FF}}^{2}}{r^{10} \mathcal{L}_{\mathcal{F}}^{2}}
+ \frac{2(\lambda + 1)}{r^{2}}
\right\}
\right]
 \mathcal{B} 
&
\notag \\
- \sqrt{32 \pi G \lambda |\mathcal{L}_{\mathcal{F}}|} f q
\left[ 
- \frac{32 \pi G f q^{2} \mathcal{L}_{\mathcal{F}}}{r^{5} \zeta^{2}(r)}
- \frac{2 f q^{2} \mathcal{L}_{\mathcal{FF}}}{r^{7} \zeta(r) \mathcal{L}_{\mathcal{F}}}
+ \frac{1}{r^{3}} \left( 1 - \frac{2(2\lambda - f + 2)}{\zeta(r)} - \frac{4 \lambda f}{\zeta^{2}(r)} \right)
\right]
\mathcal{R}^{+}
&= 0
\,.
\label{eq4-2-15}
\end{align}
We can rewrite the equations above as
\begin{align}
\frac{1}{f} \left[
\frac{1}{r} \frac{d}{dr^{*}} \left( r^{2} \frac{d}{dr^{*}} 
\left( \frac{\mathcal{R}^{+}}{r} \right) \right) + \omega^{2} \mathcal{R}^{+} 
\right] 
- V^{\mathrm{II}}_{11} \mathcal{R}^{+}
- V^{\mathrm{II}}_{12} \mathcal{B}
&= 0 \,,
\label{eq4-2-18}\\
\frac{1}{f} \left[
\frac{1}{\sqrt{|\mathcal{L}_{\mathcal{F}}|}} \frac{d}{dr^{*}} \left( \mathcal{L}_{\mathcal{F}} \frac{d}{dr^{*}} 
\left( \frac{\mathcal{B}}{\sqrt{|\mathcal{L}_{\mathcal{F}}|}} \right) \right) 
+ \mathrm{sgn}(\mathcal{L}_{\mathcal{F}}) \omega^{2} \mathcal{B}
\right] 
- V^{\mathrm{II}}_{22} \mathcal{B}
- V^{\mathrm{II}}_{21} \mathcal{R}^{+}
&= 0 \,,
\label{eq4-2-18-2}
\end{align}
where
\begin{align}
V^{\mathrm{II}}_{11} &= 
- \frac{2 \lambda}{r^{2}} + \frac{8 \lambda(\lambda - f + 1)}{r^{2} \zeta(r)}
+ \frac{8 f \lambda^{2}}{r^{2} \zeta^{2}(r)}
+ \frac{64 \pi G f \lambda q^{2} \mathcal{L}_{\mathcal{F}}}{r^{4} \zeta^{2}(r)} \,,
\label{eq4-2-19}\\
V^{\mathrm{II}}_{12} = V^{\mathrm{II}}_{21} &=
\mathrm{sgn}(\mathcal{L}_{\mathcal{F}})
\sqrt{32 \pi G \lambda |\mathcal{L}_{\mathcal{F}}| } q
\left[
- \frac{32 \pi G f q^{2} \mathcal{L}_{\mathcal{F}}}{r^{5} \zeta^{2}(r) }
- \frac{2 f q^{2} \mathcal{L}_{\mathcal{FF}}}{r^{7} \zeta(r) \mathcal{L}_{\mathcal{F}}}
+ \frac{1}{r^{3}} 
\left( 
1 - \frac{2(2\lambda - f + 2)}{\zeta(r)} - \frac{4 f \lambda}{\zeta^{2}(r)}
 \right) 
\right] \,,
\label{eq4-2-20}\\
V^{\mathrm{II}}_{22} &=
\mathrm{sgn}(\mathcal{L}_{\mathcal{F}})
\left[ 
\frac{2(\lambda + 1)}{r^{2}}
- \frac{16 \pi G q^{2} \mathcal{L}_{\mathcal{F}}}{r^{4}}
+ \frac{64 \pi G q^{2} \mathcal{L}_{\mathcal{F}}}{r^{4} \zeta (r)}
\left( \lambda + 1 + \frac{f \lambda}{\zeta(r)} + \frac{f q^{2} \mathcal{L}_{\mathcal{FF}}}{r^{4} \mathcal{L}_{\mathcal{F}}} \right)
\right.
\notag \\
&\qquad\qquad\qquad
\left.
+ \frac{2(\lambda + 1) q^{2} \mathcal{L}_{\mathcal{FF}}}{r^{6} \mathcal{L}_{\mathcal{F}}}
+ \frac{8 (8 \pi G)^{2} f q^{4} \mathcal{L}_{\mathcal{F}}^{2}}{r^{6} \zeta^{2}(r)}
\right]\,.
\label{eq4-2-21}
\end{align}
The action which yields the equations of motion for the type-II perturbations is
\begin{align}
S_{\mathrm{II}}[\mathcal{R}^{+}, \mathcal{B}] 
= \int dt dr^{*} f \, \bigg[
&- r^{2}
 \nabla^{a} \left( \frac{\mathcal{R}^{+}}{r} \right)^{*}
 \nabla_{a} \left( \frac{\mathcal{R}^{+}}{r} \right)
- \mathcal{L}_{\mathcal{F}}
 \nabla^{a} \left( \frac{\mathcal{B}}{\sqrt{|\mathcal{L}_{\mathcal{F}}|}} \right)^{*}
 \nabla_{a} \left( \frac{\mathcal{B}}{\sqrt{|\mathcal{L}_{\mathcal{F}}|}} \right)
-
\begin{pmatrix}
\mathcal{R}^{+*} & \mathcal{B}^{*}
\end{pmatrix} 
\bm{V}_{\mathrm{II}}
\begin{pmatrix}
\mathcal{R}^{+} \\
\mathcal{B}
\end{pmatrix} 
\bigg] \,,
\label{eq4-2-30}
\end{align}
where an index $a$ runs over $(t, r^{*})$, and the potential matrix is
\begin{align}
\bm{V}_{\mathrm{II}} =
\begin{pmatrix}
V^{\mathrm{II}}_{11}	&V^{\mathrm{II}}_{12} \\
\star	&V^{\mathrm{II}}_{22}
\end{pmatrix} \,.
\label{eq4-2-30-1}
\end{align}
We can read the ghost-free condition
\begin{align}
\mathcal{L}_{\mathcal{F}} > 0 \,,
\label{eq4-2-30-2}
\end{align}
which is the same as type-I, see Eq.\,\eqref{eq4-1-11}.

Given the solutions $\mathcal{R}^{+}$ and $\mathcal{B}$, we have the perturbations of the metric as
\begin{align}
K &= \frac{f}{\sqrt{2 \lambda} (\lambda + 1)} \mathcal{R}^{+\prime}
+ \frac{1}{r \sqrt{2 \lambda} (\lambda + 1) } 
\left( 1 + \lambda - f + \frac{2 f \lambda}{\zeta(r)} \right)
\mathcal{R}^{+}
- \frac{\sqrt{16 \pi G} f q \mathcal{L}_{\mathcal{F}}}{r^{2} \zeta(r) (\lambda +1) \sqrt{\mathcal{L}_{\mathcal{F}}}}
\mathcal{B} \,,
\label{eq4-2-31}\\
H_{1} &= - \frac{i \omega r}{\sqrt{2 \lambda} (\lambda + 1)}\mathcal{R}^{+\prime} 
+ \frac{i \omega}{f \sqrt{2\lambda} (\lambda +1)}
\left( \frac{\zeta(r)}{2}  -1 - \lambda + f - \frac{2 f \lambda}{\zeta(r)} \right) \mathcal{R}^{+}
+ \frac{i \omega \sqrt{16 \pi G} q \mathcal{L}_{\mathcal{F}} }{r \zeta(r) (\lambda +1) \sqrt{\mathcal{L}_{\mathcal{F}}}} \mathcal{B}
\,,
\label{eq4-2-32}\\
H_{0} =H_{2} &= -\frac{r}{\sqrt{2\lambda} (\lambda +1)}
\left( \frac{\zeta(r)}{2}  -1 - \lambda + f - \frac{2 f \lambda}{\zeta(r)} \right) 
\left( \frac{\mathcal{R}^{+}}{r} \right)'
\notag \\
&\quad + \frac{1}{\sqrt{2\lambda} (\lambda + 1) f } 
\left[
\frac{32 \pi G f^{2} \lambda q^{2} \mathcal{L}_{\mathcal{F}}}{r^{2} \zeta^{2}(r)} 
+ \frac{ f\lambda}{\zeta(r)} \left( - \zeta(r) - 2f + 4 \lambda + 4 + \frac{4 f \lambda}{\zeta(r)} \right) 
- \omega^{2} r^{2}
\right] \frac{\mathcal{R}^{+}}{r}
\notag \\
&\quad - \frac{\sqrt{16 \pi G} f q \mathcal{L}_{\mathcal{F}}}{r (\lambda + 1)\zeta(r) }
\left(  \frac{\mathcal{B}}{\sqrt{\mathcal{L}_{\mathcal{F}}}} \right)'
- \frac{\sqrt{4 \pi G} q \mathcal{L}_{\mathcal{F}}}{r^{2} (\lambda + 1)}
\left(
\frac{32 \pi G f q^{2} \mathcal{L}_{\mathcal{F}}}{r^{2} \zeta^{2}(r)} -1 + \frac{4 \lambda - 2f + 4}{\zeta(r)} + \frac{4 f \lambda}{\zeta^{2}(r)}
\right) \frac{\mathcal{B}}{\sqrt{\mathcal{L}_{\mathcal{F}}}} 
\,,
\label{eq4-2-33}
\end{align}
and the electromagnetic perturbations as
\begin{align}
f^{-}_{23} &= \frac{q}{\sqrt{2 \lambda}} \frac{\mathcal{R}^{+}}{r} + \frac{1}{\sqrt{16 \pi G }} \frac{\mathcal{B}}{\sqrt{\mathcal{L}_{\mathcal{F}}}} \,,
\label{eq4-2-35}\\
f^{-}_{02} &= - \frac{i \omega}{l(l+1)} 
\left[
\frac{q}{\sqrt{2 \lambda}} \frac{\mathcal{R}^{+}}{r} + \frac{1}{\sqrt{16 \pi G }} \frac{\mathcal{B}}{\sqrt{\mathcal{L}_{\mathcal{F}}}} 
\right]
\,,
\label{eq4-2-36}\\
f^{-}_{12} &= 
\frac{1}{l(l+1)} 
\left[
\frac{q}{\sqrt{2 \lambda}}
\left( \frac{\mathcal{R}^{+}}{r} \right)'
+ \frac{1}{\sqrt{16 \pi G }} 
\left( \frac{\mathcal{B}}{\sqrt{\mathcal{L}_{\mathcal{F}}}} \right)'
\, \right]
\,.
\label{eq4-2-37}
\end{align}


\subsubsection{Type-II master equations for $l =1$}

For the modes $l = 1$, the variable $K$ is not defined, and only the electromagnetic perturbations are dynamical.
We assume $\mathcal{L}_{\mathcal{F}} > 0$ for ghost-freeness as seen in the modes $l \geq 2$.
We define the master variable for $l = 1$ as
\begin{align}
\mathcal{B} \coloneqq \sqrt{\mathcal{L}_{\mathcal{F}}} f^{-}_{23} \,.
\label{eq4-2-2-1}
\end{align}
Then we have the master equations,
\begin{align}
\frac{d^{2} \mathcal{B}}{dr^{*2}} +
\left[ \omega^{2}
- f \left\{
\frac{2}{r^{2}} 
+ \frac{q^{2} \mathcal{L}_{\mathcal{FF}}}{r^{4} \mathcal{L}_{\mathcal{F}}}
\left( 8 \pi G \mathcal{L} + \Lambda + \frac{6f + 1}{r^{2}} \right)
+ \frac{2f q^{4} \mathcal{L}_{\mathcal{FFF}}}{r^{10} \mathcal{L}_{\mathcal{F}}}
- \frac{f q^{4} \mathcal{L}_{\mathcal{FF}}^{2}}{r^{10} \mathcal{L}_{\mathcal{F}}^{2}}
\right\}
\right]
 \mathcal{B} 
&= 0
\,,
\label{eq4-2-2-2}
\end{align}
which can be rewritten as
\begin{align}
\frac{1}{f}
\left[
\frac{1}{\sqrt{\mathcal{L}_{\mathcal{F}}}} \frac{d}{dr^{*}} \left( \mathcal{L}_{\mathcal{F}} \frac{d}{dr^{*}} 
\left( \frac{\mathcal{B}}{\sqrt{\mathcal{L}_{\mathcal{F}}}} \right) \right) 
+ \omega^{2} \mathcal{B}
\right] 
- \frac{2}{r^{2}} \left( 1 + \frac{q^{2} \mathcal{L}_{\mathcal{FF}}}{r^{4} \mathcal{L}_{\mathcal{F}}} \right) \mathcal{B}
&= 0 \,.
\label{eq4-2-2-3}
\end{align}
In terms of the master variable $\mathcal{B}$, the electromagnetic perturbations are given by
\begin{align}
f^{-}_{23} &= \frac{\mathcal{B}}{\sqrt{\mathcal{L}_{\mathcal{F}}}} \,,
\label{eq4-2-2-3-2} \\
f^{-}_{02} &= - \frac{i \omega}{2} \frac{\mathcal{B}}{\sqrt{\mathcal{L}_{\mathcal{F}}}} \,,
\label{eq4-2-2-4}\\
f^{-}_{12} &= \frac{1}{2} \left( \frac{\mathcal{B}}{\sqrt{\mathcal{L}_{\mathcal{F}}}} \right)' \,.
\label{eq4-2-2-5}
\end{align}
The metric perturbations are non-dynamical and determined by the electromagnetic perturbations:
\begin{align}
H_{2} &= \frac{r^{2} f \zeta(r)}{2q(\omega^{2}r^{2} + 1 - f )}  
\left[ \left( \frac{\mathcal{B}}{\sqrt{\mathcal{L}_{\mathcal{F}}}} \right)''
- \left(
\frac{16 \pi G q^{2} \mathcal{L}_{\mathcal{F}}}{r^{2} \zeta(r)}
-2 - \frac{\zeta(r)}{f} + \frac{2}{f} + \frac{2q^{2} \mathcal{L}_{\mathcal{FF}}}{r^{4} \mathcal{L}_{\mathcal{F}}}
\right)
\frac{1}{r}
\left( \frac{\mathcal{B}}{\sqrt{\mathcal{L}_{\mathcal{F}}}} \right)'
\right.
\notag \\
&\qquad\qquad\qquad\qquad\qquad\left. - 
\left( \frac{16 \pi G q^{2} \mathcal{L}_{\mathcal{F}}}{r^{4} f \zeta(r) } 
- \frac{\omega^{2}}{f^{2}} + \frac{2}{r^{2}f} \left( 1 + \frac{q^{2} \mathcal{L}_{\mathcal{FF}}}{r^{4} \mathcal{L}_{\mathcal{F}}} \right) 
\right) \frac{\mathcal{B}}{\sqrt{\mathcal{L}_{\mathcal{F}}}}
\right] \,,
\label{eq4-2-2-6}\\
H_{1} &= \frac{i \omega r^{3} f \zeta(r)}{2q(\omega^{2}r^{2} + 1 - f )}  
\left[ \left( \frac{\mathcal{B}}{\sqrt{\mathcal{L}_{\mathcal{F}}}} \right)''
- \left(
\frac{16 \pi G q^{2} \mathcal{L}_{\mathcal{F}}}{r^{2} \zeta(r)}
-2 - \frac{\zeta(r)}{f} + \frac{2}{f} + \frac{2q^{2} \mathcal{L}_{\mathcal{FF}}}{r^{4} \mathcal{L}_{\mathcal{F}}}
\right)
\frac{1}{r}
\left( \frac{\mathcal{B}}{\sqrt{\mathcal{L}_{\mathcal{F}}}} \right)'
\right.
\notag \\
&\qquad\qquad\qquad\qquad\qquad\left. - 
\left( \frac{16 \pi G q^{2} \mathcal{L}_{\mathcal{F}}}{r^{4} f \zeta(r) } 
- \frac{\omega^{2}}{f^{2}} + \frac{2}{r^{2}f} \left( 1 + \frac{q^{2} \mathcal{L}_{\mathcal{FF}}}{r^{4} \mathcal{L}_{\mathcal{F}}} \right) 
\right) \frac{\mathcal{B}}{\sqrt{\mathcal{L}_{\mathcal{F}}}}
\right] \,,
\label{eq4-2-2-7}\\
H_{0} &= - \frac{r^{2}f (2 \omega^{2} r^{2} + 2 - 2f - \zeta(r))}{2q(\omega^{2} r^{2} + 1 - f)} 
\notag \\
&\qquad \times
\left[
\left( \frac{\mathcal{B}}{\sqrt{\mathcal{L}_{\mathcal{F}}}} \right)''
+
\left( \frac{16 \pi G q^{2} \mathcal{L}_{\mathcal{F}}}{r^{2}(2\omega^{2} r^{2} +2 -2f - \zeta(r))} + 2 + \frac{\zeta(r)}{f} - \frac{2}{f} -\frac{2q^{2} \mathcal{L}_{\mathcal{FF}}}{r^{4} \mathcal{L}_{\mathcal{F}}} \right)
\frac{1}{r}
\left( \frac{\mathcal{B}}{\sqrt{\mathcal{L}_{\mathcal{F}}}} \right)'
\right.
\notag \\
&\qquad\qquad \left. +
\left(\frac{16 \pi G q^{2} \mathcal{L}_{\mathcal{F}}}{r^{4} f (2\omega^{2} r^{2} +2 -2f - \zeta(r))} +  \frac{\omega^{2}}{f^{2}}  - \frac{2}{r^{2}f}  \left( 1 + \frac{q^{2} \mathcal{L}_{\mathcal{FF}}}{r^{4} \mathcal{L}_{\mathcal{F}}} \right)  \right) \frac{\mathcal{B}}{\sqrt{\mathcal{L}_{\mathcal{F}}}} 
\right] \,.
\label{eq4-2-2-8}
\end{align}

\section{Stability conditions}
\label{sec:stability}

In this section, we derive sufficient conditions for the stability of magnetically charged black holes in general nonlinear electrodynamics.
The procedure here basically follows Ref.\,\cite{Moreno:2002gg}.
The background solutions are stable against linear perturbations for the modes $l \geq 2$ if the potential matrices $\bm{V}_{\mathrm{I}}$ and $\bm{V}_{\mathrm{II}}$ are positive-definite.
For a $2 \times 2$ matrix, it is equivalent to that both the determinant and trace of the matrix are positive.
Thus we first calculate the determinant and trace of the potential matrices, and propose the sufficient conditions for both of them to be positive.
We derive the stability conditions for the type-I and type-II perturbations separately.
Finally, we summarize them.

\subsection{Stability condition for type-I perturbations}

For the type-I potential $\bm{V}_{\mathrm{I}}$ defined for $l \geq 2$, we have the determinant and trace as 
\begin{align}
\det (\bm{V}_{\mathrm{I}}) &= \frac{(l-1)l(l+1)(l+2)}{r^{4}} \frac{1}{1 - q^{2} \mathcal{L}_{\mathcal{\widetilde{F}\widetilde{F}}} / (r^{4} \mathcal{L}_{\mathcal{F}}) }
 \,,
\label{eq4-3-2}\\
\mathrm{tr} (\bm{V}_{\mathrm{I}}) &=
\frac{(l+2)(l-1)}{r^{2}} + \frac{16 \pi G q^{2} \mathcal{L}_{\mathcal{F}}}{r^{4}} + \frac{l(l+1)}{r^{2}} \frac{1}{1 - q^{2} \mathcal{L}_{\mathcal{\widetilde{F}\widetilde{F}}} / (r^{4} \mathcal{L}_{\mathcal{F}} )}
\,.
\label{eq4-3-3}
\end{align}
We can read off the stability condition of the background solution against the type-I perturbations as
\begin{align}
1 - \frac{q^{2} \mathcal{L}_{\mathcal{\widetilde{F}\widetilde{F}}}(\bar{\mathcal{F}}, 0) }{r^{4} \mathcal{L}_{\mathcal{F}}(\bar{\mathcal{F}}, 0)}
= 1 - 2 \bar{\mathcal{F}}  \frac{ \mathcal{L}_{\mathcal{\widetilde{F}\widetilde{F}}}(\bar{\mathcal{F}}, 0) }{ \mathcal{L}_{\mathcal{F}}(\bar{\mathcal{F}}, 0) }
> 0 \,.
\label{eq4-3-4}
\end{align}
Note that $\mathcal{L}_{\mathcal{F}}$ must be positive for ghost-free theory, see Eq.\,\eqref{eq4-1-11}.
When the condition is satisfied, the stability against the electromagnetic perturbations of the modes $l = 1$ also holds as we can see from Eq.\,\eqref{eq4-1-2-3}.

\subsection{Stability condition for type-II perturbations}

For the type-II potential, we have
\begin{align}
\det (\bm{V}_{\mathrm{II}}) &=
\frac{128 \pi G \lambda f q^{2} \mathcal{L}_{\mathcal{F}}}{r^{6} \zeta^{2}(r)} \left( 1 + \frac{q^{2} \mathcal{L}_{\mathcal{FF}}}{r^{4} \mathcal{L}_{\mathcal{F}}} \right) 
\left[ \lambda + 1 - f \left( 1 + \frac{q^{2} \mathcal{L}_{\mathcal{FF}}}{r^{4} \mathcal{L}_{\mathcal{F}}} \right) \right]
\notag \\
&\quad +
\frac{4 \lambda (\lambda +1)}{r^{4} \zeta^{2}(r)} \left( 1 + \frac{q^{2} \mathcal{L}_{\mathcal{FF}}}{r^{4} \mathcal{L}_{\mathcal{F}}} \right) 
\left[ (- \zeta(r) - 4f + 4\lambda + 4) \zeta(r)  + 4 \lambda f \right] \,,
\label{eq4-3-5}\\
\mathrm{tr} (\bm{V}_{\mathrm{II}}) &=
\frac{2}{r^{2}}
\left(
\lambda + 1
+ \frac{32\pi G f q^{2} \mathcal{L}_{\mathcal{F}}}{r^{2} \zeta(r)} 
\right)
\left( 1 + \frac{q^{2} \mathcal{L}_{\mathcal{FF}}}{r^{4} \mathcal{L}_{\mathcal{F}}} \right)
+ \frac{2f}{r^{2} \zeta^{2}(r)} \left( 2\lambda + \frac{16 \pi G q^{2} \mathcal{L}_{\mathcal{F}}}{r^{2}} \right)^{2} 
\notag \\
&\quad+ \frac{- \zeta(r) - 4f + 4\lambda + 4}{r^{2} \zeta(r)} 
\left( 2\lambda + \frac{16 \pi G q^{2} \mathcal{L}_{\mathcal{F}}}{r^{2}} \right) \,.
\label{eq4-3-6}
\end{align}
Below, we derive the sufficient condition for stability under the following assumptions,
\begin{align}
\mathcal{L}(\bar{\mathcal{F}}, 0) + \frac{\Lambda}{8 \pi G} > 0 \quad \text{and} \quad
1 + \frac{q^{2} \mathcal{L}_{\mathcal{FF}}(\bar{\mathcal{F}}, 0) }{r^{4} \mathcal{L}_{\mathcal{F}}(\bar{\mathcal{F}}, 0)}
= 1 + 2 \bar{\mathcal{F}} \frac{ \mathcal{L}_{\mathcal{FF}}(\bar{\mathcal{F}}, 0) }{\mathcal{L}_{\mathcal{F}}(\bar{\mathcal{F}}, 0)}
 > 0 \,.
\label{eq4-3-7}
\end{align}
The former is reasonable as the weak energy condition implies, see Eq.\,\eqref{eq2-53}.
First, let us show that the function $\zeta (r)$ defined by Eq.\,\eqref{eq4-2-17} is positive outside the horizon of the black hole.
Let $r = r_{h}$ be the position of the horizon, where $f(r_{h}) = 0$ and $f(r)$ becomes from negative to positive from inside to outside the horizon smoothly.
We therefore have $\zeta(r_{h}) = r_{h} f'(r_{h}) + l(l+1) > 0$.
In addition, a brief calculation gives $(r \zeta(r))' = 16 \pi G q^{2} \mathcal{L}_{\mathcal{F}} / r^{2} + (l+2)(l-1) > 0$ for $l \geq 1$, where we used the background equations \eqref{eq2-13} and \eqref{eq2-13-2}, and the ghost-free condition $\mathcal{L}_{\mathcal{F}} > 0$.
Thus it is shown that $\zeta(r)$ is positive outside the horizon.
Next, note that it follows $0 < f(r > r_{h}) < 1$ from Eq.\,\eqref{eq2-14} with a positive constant $M$.
Then using the fact that $\zeta(r > r_{h}) > 0$ and $0 < f(r > r_{h}) < 1$, we can see that $- \zeta(r) - 4f + 4\lambda + 4 = r^{2}(8 \pi G \mathcal{L} + \Lambda) + 1- f + 2\lambda > 0$ outside the horizon.
Therefore, $\mathrm{tr}(\bm{V}_{\mathrm{II}})$ is positive outside the horizon under the assumptions \eqref{eq4-3-7}.
On the other hand, $\det(\bm{V}_{\mathrm{II}})$ is always positive outside the horizon if
\begin{align}
f \left( 1 + \frac{q^{2} \mathcal{L}_{\mathcal{FF}}(\bar{\mathcal{F}}, 0)}{r^{4} \mathcal{L}_{\mathcal{F}} (\bar{\mathcal{F}}, 0)} \right) 
= f \left( 1 + 2 \bar{\mathcal{F}} \frac{\mathcal{L}_{\mathcal{FF}}(\bar{\mathcal{F}}, 0)}{\mathcal{L}_{\mathcal{F}} (\bar{\mathcal{F}}, 0)} \right)
< \lambda + 1 = \frac{l(l+1)}{2} \,.
\label{eq4-3-8}
\end{align}
Under the assumptions \eqref{eq4-3-7}, this relation gives the sufficient condition for the background solution to be stable with respect to the type-II perturbations for the modes $l \geq 2$.
For $l = 1$, if we assume the latter of Eq.\,\eqref{eq4-3-7}, the stability is ensured as we can see from Eq.\,\eqref{eq4-2-2-3}.

\subsection{Summary of stability conditions}

For later convenience, let us summarize the sufficient conditions for the stability of magnetic
 black holes:
\begin{align}
\mathcal{L}(\bar{\mathcal{F}}, 0) + \frac{\Lambda}{8 \pi G} > 0 \,,
\label{eq4.2-C-1}\\
\mathcal{L}_{\mathcal{F}} (\bar{\mathcal{F}}, 0) > 0 \,,
\label{eq4.2-C-2}\\
1 - 2 \bar{\mathcal{F}}  \frac{ \mathcal{L}_{\mathcal{\widetilde{F}\widetilde{F}}}(\bar{\mathcal{F}}, 0) }{ \mathcal{L}_{\mathcal{F}}(\bar{\mathcal{F}}, 0) }
> 0 \,,
\label{eq4.2-C-3}\\
0 < f \left( 1 + 2 \bar{\mathcal{F}} \frac{\mathcal{L}_{\mathcal{FF}}(\bar{\mathcal{F}}, 0)}{\mathcal{L}_{\mathcal{F}} (\bar{\mathcal{F}}, 0)} \right)
< 3 \,,
\label{eq4.2-C-4}
\end{align} 
where the upper bound in the condition \eqref{eq4.2-C-4} comes from setting $l=2$ in Eq.\,\eqref{eq4-3-8}.
Notice that the conditions \eqref{eq4.2-C-1} and \eqref{eq4.2-C-2} together form the weak energy condition \eqref{eq2-53}.
The conditions \eqref{eq4.2-C-1}, \eqref{eq4.2-C-2} and \eqref{eq4.2-C-4} have been already derived in Ref.\,\cite{Moreno:2002gg}, while the condition \eqref{eq4.2-C-3} is a new result obtained
 by taking into account $\widetilde{\mathcal{F}}$ dependence of the theory.

\section{Applications}
\label{sec:applications}

In this section, we apply the stability conditions derived in the previous section to specific models.
Since we focus on the asymptotically flat spacetime, hereafter we set the cosmological constant $\Lambda$ to zero. 
Note that we are considering the magnetically charged background, and the invariants of the background electromagnetic field are given by Eq.\,\eqref{eq4-5},
\begin{align}
\bar{\mathcal{F}} = \frac{q^{2}}{2r^{4}}\,, \qquad
\bar{\mathcal{\widetilde{F}}} = 0 \, .
\label{eq.m5.1}
\end{align}

%

\subsection{Bardeen-like regular black holes}
\label{subsec:BardeenBH}

Let us choose the Lagrangian as
\begin{align}
\mathcal{L}(\mathcal{F}, \widetilde{\mathcal{F}})
= \mu^{4}
\left( \frac{\sqrt{ \mathcal{F} / \mu^{4}}}{1 + \gamma \sqrt{\mathcal{F} / \mu^{4}}} \right)^{N} + \sum_{n=2}^{\infty} \frac{1}{n!} \mathcal{L}_{n}(\mathcal{F}) \widetilde{\mathcal{F}}^{n}
\, ,
\label{eq4-B-1}
\end{align}
where $\mu$ is a parameter with a mass dimension, $\gamma$ and $N$ are dimensionless non-negative constants, and $\mathcal{L}_{n}(\mathcal{F})$ are  functions of $\mathcal{F}$.
We choose the form of the first term as an extension of the model in
Ref.\,\cite{Chaverra:2016ttw}, which resolves a singularity at the center of  black holes, as we will see below.
Here we add the terms proportional to $\widetilde{\mathcal{F}}$ like Eq.\,\eqref{eq2-17}.
From Eq.\,\eqref{eq2-14}, we obtain the metric function $f(r)$ as 
\begin{align}
f(r) = 1 - \frac{2GM}{r} + \frac{8 \pi G \cdot 2^{-N/2} r^{2-2N} \mu^{4-2N} |q|^{N}}{2N-3}
\, {}_{2}F_{1} \left( - \frac{3}{2} + N, N, - \frac{1}{2} + N, - \frac{\gamma |q|}{\sqrt{2} r^{2} \mu^{2}} \right) \,,
\label{eq4-B-2}
\end{align} 
where ${}_{2}F_{1}(\cdot , \cdot , \cdot , \cdot)$ is the hypergeometric function.
The radial position of the horizon is determined by a zero of $f(r)$.
Note that the weak energy condition \eqref{eq2-53} is satisfied since $\mathcal{L}(\bar{\mathcal{F}}, 0)$ and 
\begin{align}
\mathcal{L}_{\mathcal{F}}(\bar{\mathcal{F}}, 0) &= \frac{\mu^{4} N}{2 \bar{\mathcal{F}}} \frac{(\sqrt{\bar{\mathcal{F}}/\mu^{4}})^{N}}{(1 + \gamma \sqrt{\bar{\mathcal{F}}/\mu^{4}})^{N+1}}
\label{eq4-B-2-2}
\end{align} 
are always positive.
On the other hand, in general, the strong energy condition which is an assumption for Penrose--Hawking singularity theorem can be violated in a particular region.
We here list the quantities which have importance for the stability analysis,
\begin{align}
1 - 2 \bar{\mathcal{F}} \frac{ \mathcal{L}_{\mathcal{\widetilde{F}\widetilde{F}}} (\bar{\mathcal{F}}, 0) }{\mathcal{L}_{\mathcal{F}}(\bar{\mathcal{F}}, 0)}
&= 1- \frac{4 \bar{\mathcal{F}}}{N} 
\frac{(1+ \gamma \sqrt{\bar{\mathcal{F}} / \mu^{4} })^{N+1}}{ ( \sqrt{\bar{\mathcal{F}} / \mu^{4} } )^{N-2}} \mathcal{L}_{2}(\bar{\mathcal{F}}) \,,
\label{eq4-B-2-3}\\
1 + 2 \bar{\mathcal{F}} \frac{ \mathcal{L}_{\mathcal{FF}} (\bar{\mathcal{F}}, 0) }{\mathcal{L}_{\mathcal{F}}(\bar{\mathcal{F}}, 0)} 
&= \frac{N - 1 - 2 \gamma \sqrt{\bar{\mathcal{F}} / \mu^{4} }}{ 1 + \gamma \sqrt{\bar{\mathcal{F}} / \mu^{4} } } \,.
\label{eq4-B-2-4}
\end{align} 
From Eq.\,\eqref{eq4.2-C-3}, the function $\mathcal{L}_{2}(\mathcal{F})$, which characterizes a $\widetilde{\mathcal{F}}$ dependence of the Lagrangian, does not interfere the stability against the type-I perturbations unless it makes Eq.\,\eqref{eq4-B-2-3} negative outside the horizon. In other words, the positivity of Eq.\,\eqref{eq4-B-2-3} can be understood as a condition on $\mathcal{L}_{2}(\mathcal{F})$:
\begin{eqnarray}
 \mathcal{L}_{2}(\bar{\mathcal{F}})  \leq    \frac{N} {4 \bar{\mathcal{F}}}
\frac{ ( \sqrt{\bar{\mathcal{F}} / \mu^{4} } )^{N-2}}{(1+ \gamma \sqrt{\bar{\mathcal{F}} / \mu^{4} })^{N+1}}
\end{eqnarray}
Next, let us focus on the stability against the type-II perturbations, which can be discussed by analyzing Eq.\,\eqref{eq4-B-2-4}.
Note that Eq.\,\eqref{eq4-B-2-4} monotonously grows as $\bar{\mathcal{F}}$ decreases and it approaches the maximum given by $N - 1$ when $\bar{\mathcal{F}} \to 0$, or equivalently, $r \to \infty$.
Thus if $N < 4$, the latter inequality in Eq.\,\eqref{eq4.2-C-4} is always satisfied.
In the following paragraphs, we investigate whether the function \eqref{eq4-B-2-4} is positive outside the horizon for some models which satisfy $N < 4$ and yield black hole solutions without curvature singularity.

Ay\'on-Beato and Garcia \cite{AyonBeato:2000zs} point out that setting $N= 5/2$ and an appropriate choice of the parameters can give rise to the so-called Bardeen's regular black hole solution.
In fact, when we choose $N = 5/2$ and 
the integration constant $M$ as
\begin{align}
M = \frac{2^{5/4} \pi \mu |q|^{3/2}}{3 \gamma} \,,
\label{eq4-B-3}
\end{align} 
the metric function \eqref{eq4-B-2} reduces to
\begin{align}
f(r) = 1 - \frac{6 \sqrt{3} G M^{5/2} \mu^{3/2} r^{2} }{(3M\mu r^{2} + 2^{3/4} \pi |q|^{5/2})^{3/2}} \,,
\label{eq4-B-4}
\end{align} 
which no longer has any divergence of the curvature invariants at $r=0$. We note that there is still monopole singularity at $r = 0$ simply because $\bar{\mathcal{F}} = q^2/2r^4$.
The metric function has a minimum at $r_{min} \coloneqq 2^{7/8} \sqrt{\pi} |q|^{5/4} / \sqrt{3 M \mu}$, and the radial position of the horizon denoted by $r_{h}$ satisfies $r_{h} > r_{min}$ if it exists.
We can see that Eq.\,\eqref{eq4-B-2-4} is always positive at $r > r_{h}$ because it is already positive at $r = r_{min}$, where its value is $1/3$.
Thus the Bardeen black hole solution proposed in Ref.\,\cite{AyonBeato:2000zs} in the context of nonlinear electrodynamics is stable
 as long as the function $\mathcal{L}_{2}(\mathcal{F})$ guarantees the positivity of Eq.\,\eqref{eq4-B-2-3} with $N = 5/2$. This result is consistent with that
in Ref.\,\cite{Moreno:2002gg},
which corresponds to the case $\mathcal{L}_{2}(\mathcal{F}) = 0$ in our setup.

\begin{figure}[t]
\centering
  \includegraphics[width=8cm]{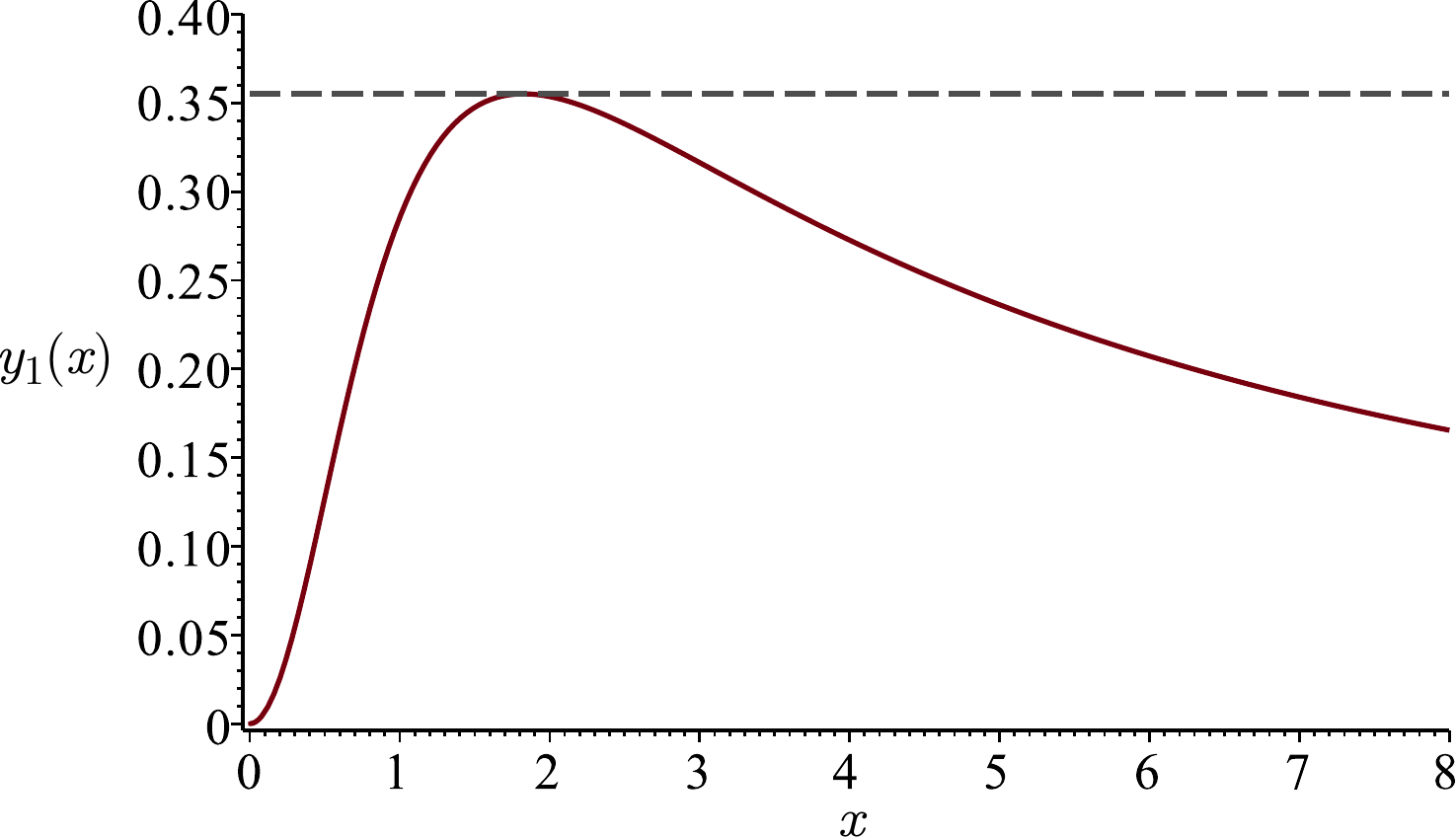}
  \caption{
  A red line indicates the behavior of $y_{1}(x)$ defined by Eq.\,\eqref{eq4-B-8-2}.
  A dashed gray line at $0.3551$ divides whether the horizons exist or not.
  }
  \label{fig1}
\end{figure}

An alternative model for the Bardeen-like regular black hole with a magnetic charge is considered in Ref.\,\cite{Chaverra:2016ttw}, which corresponds to $N = 2$.
In this case, by neglecting the latter terms in Eq.\,\eqref{eq4-B-1} and taking the weak field limit  $\gamma \sqrt{\mathcal{F} / \mu^{4}} \ll 1$, we can recover the Maxwell theory given by $\mathcal{L}(\mathcal{F}, \widetilde{\mathcal{F}}) = \mathcal{F}$.
With a particular choice of the 
integration constant,
\begin{align}
M = \frac{\pi^{2} \mu |q|^{3/2} }{2^{3/4} \sqrt{\gamma}} \,,
\label{eq4-B-5}
\end{align}
the metric function \eqref{eq4-B-2} reduces to
\begin{align}
f(r) &= 1 - \frac{2GM}{r} + \frac{8 \pi G q^{2} M^{2}}{\pi^{4} q^{4} + 4 M^{2} r^{2}} + \frac{4GM \arctan(\pi^{2} q^{2} /(2 M r))}{ \pi r }
\notag \\
&= 1 - \frac{8 G M^{2}}{\pi^{3} q^{2}} \left( \frac{\arctan x}{x} - \frac{1}{1 + x^{2}} \right)
 \,,
\label{eq4-B-6}
\end{align} 
where we defined a new variable $x \coloneqq 2Mr / (\pi^{2} q^{2})$ in the second line.
This metric is also regular at $r=0$, and asymptotically behaves like the Reissner--Nordstr\"om black hole,
\begin{align}
f(r) \xrightarrow{r \to \infty} 1 - \frac{2GM}{r} + \frac{4 \pi Gq^{2}}{r^{2}} + \mathcal{O}(r^{-3}) \,.
\label{eq4-B-7}
\end{align}
We are interested in whether Eq.\,\eqref{eq4-B-2-4} is positive everywhere outside the horizon.
Now it is given by
\begin{align}
1 + 2 \bar{\mathcal{F}} \frac{ \mathcal{L}_{\mathcal{FF}} (\bar{\mathcal{F}}, 0) }{\mathcal{L}_{\mathcal{F}}(\bar{\mathcal{F}}, 0)} 
&= \frac{ 4 M^{2} r^{2}- 2 \pi^{4} q^{4}}{4 M^{2} r^{2} + \pi^{4} q^{4} }
= \frac{x^{2} - 2}{x^{2} + 1} 
\eqqcolon Z(x) \,.
\label{eq4-B-8}
\end{align}
This function has a single zero at $x = x_{0} \coloneqq \sqrt{2}$, outside which it is always positive.
Thus our task is to verify that $x_{0} = \sqrt{2}$ is inside the horizon.
If it is true, the stability is guaranteed since outside the horizon $Z(x)$ is positive.
In order to seek the position of the horizon, that is, zero of $f(r)$ given by Eq.\,\eqref{eq4-B-6}, we look at the intersections of the function 
\begin{align}
y_{1}(x) \coloneqq \frac{\arctan x}{x} - \frac{1}{1 + x^{2}}
\label{eq4-B-8-2}
\end{align}
and $y_{2} \coloneqq \pi^{3} q^{2} / (8 G M^{2})$.
The behavior of $y_{1}(x)$ is shown in Fig.\,\ref{fig1}, from which we can see that there are two horizons when $y_{2} < 0.3551$. 
In the extremal case given by $y_{2} = 0.3551$, there is single horizon. 
When $y_{2} > 0.3551$, $f(r)$ is always positive over the spacetime and the metric describes some kind of soliton rather than black hole.
Moreover, Fig.\,\ref{fig1} tells us that the innermost position of the outer horizon is given by $x = 1.825$. 
Therefore, we can conclude that $x_{0} = \sqrt{2} < 1.825$ is always located inside the horizon if the black hole exists, and the Bardeen-like regular black hole is stable as long as the function $\mathcal{L}_{2}(\mathcal{F})$ guarantees the positivity of Eq.\,\eqref{eq4-B-2-3} with $N=2$.

%



\subsection{Euler--Heisenberg theory}

\begin{figure}[t]
\centering
  \includegraphics[width=7cm]{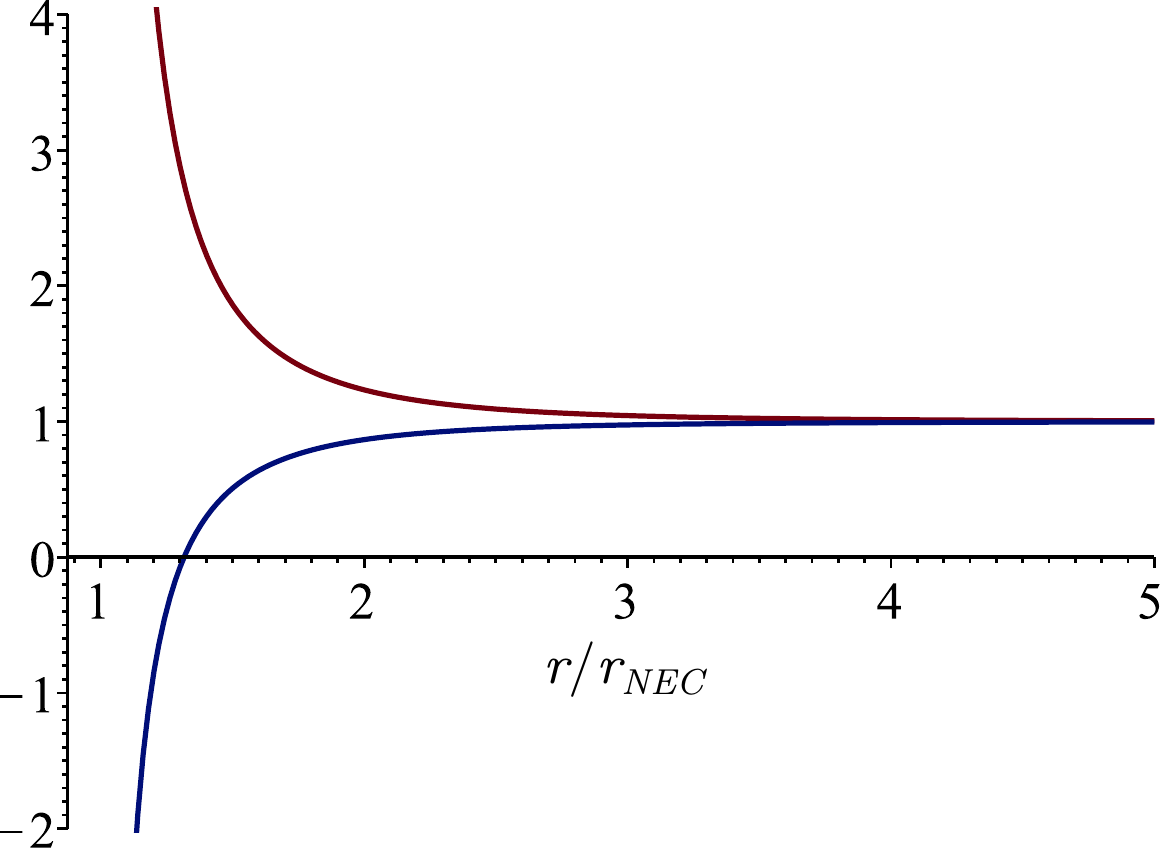}
  \caption{
  A red and a blue line indicate the behaviors of the functions \eqref{eq4-EH-4} and \eqref{eq4-EH-5}, respectively.
  }
  \label{fig2}
\end{figure}

The Euler--Heisenberg Lagrangian was first proposed in 1936 \cite{Heisenberg:1935qt}, which describes 
the effective Lagrangian for a constant electromagnetic field in the quantum electrodynamics after integrating out the electron.
In a non-perturbative form, it is given by
\begin{align}
\mathcal{L}(\mathcal{F}, \widetilde{\mathcal{F}}) = \mathcal{F} + \frac{1}{8 \pi^{2}} 
\int_{0}^{\infty} \frac{ds}{s^{3}} \exp(- m_{e}^{2} s)
\left[
(es)^{2} \frac{\textrm{Re} \cosh \left( es \sqrt{2 \mathcal{F} + 2 i \widetilde{\mathcal{F}}} \right)}{\textrm{Im} \cosh \left( es \sqrt{2 \mathcal{F} + 2 i \widetilde{\mathcal{F}}} \right)} \widetilde{\mathcal{F}} - \frac{2}{3} (es)^{2} \mathcal{F} - 1
\right] \,,
\label{eq4-EH-1}
\end{align}
where $m_{e}$ is the electron mass, and $e$ is the elementary charge. 
Assuming that the electromagnetic field is sufficiently small, and taking into account the terms up to the quadratic order of $\mathcal{F}$ and $\widetilde{\mathcal{F}}$, the Lagrangian is approximated by
\begin{align}
\mathcal{L}(\mathcal{F}, \widetilde{\mathcal{F}}) = \mathcal{F} - \frac{2 \alpha^{2}}{45 m_{e}^{4}} ( 4 \mathcal{F}^{2} + 7 \widetilde{\mathcal{F}}^{2} ) \,,
\label{eq4-EH-2}
\end{align}
where $\alpha \coloneqq e^{2} / 4 \pi$ is the fine-structure constant.
The shadow of a black hole in this quadratic theory is studied in Ref.\,\cite{Allahyari:2019jqz}.
Here let us see a stability of a black hole in this theory.
For the effective Lagrangian, we have the metric function as
\begin{align}
f(r) = 1 - \frac{2GM}{r} + \frac{4\pi G q^{2}}{r^{2}} - \frac{16 \pi G \alpha^{2} q^{4}}{225 m_{e}^{4} r^{6}} \,.
\label{eq4-EH-2-2}
\end{align}
Let us examine the stability on the effective Lagrangian briefly. 
The first derivative with respect to $\mathcal{F}$ is
\begin{align}
\mathcal{L}_{\mathcal{F}} (\bar{\mathcal{F}}, 0)
&= 1 - \frac{8 \alpha^{2} q^{2}}{45 m_{e}^{4} r^{4}} \,.
\label{eq4-EH-3}
\end{align}
We need $\mathcal{L}_{\mathcal{F}}(\bar{\mathcal{F}}, 0) > 0$ in order to satisfy the null energy condition and prevent the existence of a ghost, which is violated inside $r_{NEC} \coloneqq (8/45)^{1/4} \sqrt{\alpha |q|}/m_{e}$.
In terms of the radius, we have 
\begin{align}
1 - 2 \bar{\mathcal{F}} \frac{ \mathcal{L}_{\mathcal{\widetilde{F}\widetilde{F}}} (\bar{\mathcal{F}}, 0) }{\mathcal{L}_{\mathcal{F}}(\bar{\mathcal{F}}, 0)}
&= 1 + \frac{7}{2} \cdot \frac{1}{(r/r_{NEC})^{4} -1} \,,
\label{eq4-EH-4}\\
1 + 2 \bar{\mathcal{F}} \frac{ \mathcal{L}_{\mathcal{FF}} (\bar{\mathcal{F}}, 0) }{\mathcal{L}_{\mathcal{F}}(\bar{\mathcal{F}}, 0)} 
&= 1 - \frac{2}{(r/r_{NEC})^{4} - 1}
\,.
\label{eq4-EH-5}
\end{align} 
The behaviors of these functions are shown in Fig.\,\ref{fig2}.
We can see the function \eqref{eq4-EH-4} is always positive at $r > r_{NEC}$.
Meanwhile, the function \eqref{eq4-EH-5} is negative in the region $1 < r/r_{NEC} < 3^{1/4} = 1.3161$.
The stability depends on he mass and charge of the black hole.
Black holes are stable if  the horizon is located outside of the position where the stability condition is violated.
To see this, we evaluate the metric function at the positions of our interest,
\begin{align}
f(r_{NEC}) &= 
1 - \frac{ \sqrt{3} \,  10^{1/4} G M m_{e}}{\sqrt{|q| \alpha}}
+ \frac{27 }{8 \sqrt{10}} \frac{8 \pi G m_{e}^{2} |q|}{\alpha} \,,
\label{eq-EH-6}\\
f(3^{1/4} r_{NEC}) &= 
1 - \frac{ 30^{1/4} G M m_{e}}{\sqrt{|q| \alpha}}
+ \frac{29}{8\sqrt{30}} \frac{8 \pi G m_{e}^{2} |q|}{\alpha} \,.
\label{eq-EH-7}
\end{align}
Then it turns out that the functions \eqref{eq4-EH-3} and \eqref{eq4-EH-5} are always positive outside the horizon and the stability is ensured if
\begin{align}
\frac{\sqrt{\alpha |q|}}{30^{1/4} G m_{e}} + \frac{29 \pi m_{e} |q|^{3/2}}{30^{3/4}\sqrt{\alpha}} < M
\quad \text{and} \quad
\frac{10^{3/4} \sqrt{3 \alpha |q|}}{30 G m_{e}} + \frac{9 \sqrt{3} \pi m_{e} |q|^{3/2}}{10^{3/4} \sqrt{\alpha}}
< M \,.
\label{eq-EH-8}
\end{align} 
On the other hand, the black hole can be unstable if the charge is sufficiently large compared to the mass.
Note that this instability is a prediction of the {\it quadratic} Euler--Heisenberg Lagrangian \eqref{eq4-EH-2}, not the quantum electrodynamics itself because of following two reasons: First, $r_{NEC}$ is nothing but a length scale where $\bar{{\cal F}} = q^2/ 2r^4 \sim m_{e}^4/ \alpha^2$. Thus the truncation of the higher order terms is no longer valid near $r \sim r_{NEC}$. 
Second, the spherically symmetric configuration means that the electromagnetic field is not a constant. In such a case, non-perturbative Euler--Heisenberg Lagrangian is no longer valid and derivative interactions such as $\partial_{\mu} F_{\nu\rho} \partial^{\mu} F^{\nu\rho}$ will contribute to the effective Lagrangian.
Thus, the appearance of instability in our analysis should be understood as the necessity of such an ultraviolet completion of the quadratic theory.

\subsection{Born--Infeld theory}

In the 1930s, Born and Infeld introduced a nonlinear Lagrangian to remove the divergence of the electron's self-energy in classical electrodynamics \cite{Born:1934gh}.
Black hole solutions in a class of Born--Infeld theory have been studied in Refs.\,\cite{Demianski:1986wx, Breton:2003tk, Breton:2007bza, Kruglov:2017mpj}.
The Lagrangian is given by
\begin{align}
\mathcal{L}(\mathcal{F}, \widetilde{\mathcal{F}})
= \mu^{4} \sqrt{ 1 + \frac{2 \mathcal{F}}{\mu^{4}} - \frac{\widetilde{\mathcal{F}}^{2}}{\mu^{8}} }
- \mu^{4} \,,
\label{eq4-C-1}
\end{align} 
where $\mu$ is a scale parameter with a mass dimension.
When the electromagnetic field is sufficiently small compared to the scale, the Lagrangian reduces to Maxwell's one.
For the Lagrangian, we obtain the metric function as
\begin{align}
f(r) = 1 - \frac{2GM}{r} + \frac{8 \pi G \mu^{4} r^{2}}{3} \left[ 1 - {}_{2}F_{1}\left( - \frac{3}{4}, - \frac{1}{2}, \frac{1}{4}, - \frac{q^{2}}{\mu^{4} r^{4}} \right) \right] \,.
\label{eq4-C-1-2}
\end{align} 
For the stability analysis, we need the following expressions
\begin{align}
\mathcal{L}_{\mathcal{F}} (\bar{\mathcal{F}}, 0)
&= \frac{1 }{\sqrt{ 1 + 2 \bar{\mathcal{F}} / \mu^{4} } } \,,
\label{eq4-C-2}\\
1 - 2 \bar{\mathcal{F}} \frac{ \mathcal{L}_{\mathcal{\widetilde{F}\widetilde{F}}} (\bar{\mathcal{F}}, 0) }{\mathcal{L}_{\mathcal{F}}(\bar{\mathcal{F}}, 0)}
&= 1 + \frac{2 \bar{\mathcal{F}}}{\mu^{4}} \,,
\label{eq4-C-3}\\
1 + 2 \bar{\mathcal{F}} \frac{ \mathcal{L}_{\mathcal{FF}} (\bar{\mathcal{F}}, 0) }{\mathcal{L}_{\mathcal{F}}(\bar{\mathcal{F}}, 0)} 
&= \frac{1}{1 + 2 \bar{\mathcal{F}} / \mu^{4} } \,,
\label{eq4-C-4}
\end{align} 
which are all positive.
In addition, the quantity \eqref{eq4-C-4} is less than one.
Thus the stability conditions derived in the previous section are all satisfied in this theory.

\section{Summary and Outlook}
\label{sec:discussion}

In this paper, we obtained equations of motion for the linear perturbations on the magnetically charged black hole in general nonlinear electrodynamics.
Furthermore, we clarified sufficient conditions for the stability of {magnetic} black holes, which are summarized by Eqs.\,\eqref{eq4.2-C-1}--\eqref{eq4.2-C-4}. Comparing to the previous work {for electric black holes} by Moreno and Sarbach~\cite{Moreno:2002gg}, where they focus on a Lagrangian with $\mathcal L(\mathcal{F})$, we investigated the most general action \eqref{eq2-0-1} that is composed of the field strength $F_{\mu\nu}$ and its Hodge dual $\widetilde{F}_{\mu\nu}$, which generally reduces to a Lagrangian with $\mathcal{L}(\mathcal{F}, \widetilde{\mathcal{F}})$ given by Eq.\,\eqref{eq2-1} as demonstrated in Appendix \ref{app:General form}.
We checked the stability conditions in three models {with a magnetic charge}: Bardeen's regular black holes, black holes in Euler--Heisenberg theory and black holes in Born--Infeld theory. 
We obtained a sufficient condition of $\widetilde{{\cal F}}$ dependence for Bardeen's {magnetic} black holes, which is the positivity of Eq.\,\eqref{eq4-B-2-3}.  We showed that {magnetic} black holes in the quadratic Euler--Heisenberg theory are stable when the mass and the charge satisfy the condition \eqref{eq-EH-8}.
{In addition, we proved that magnetic black holes in Born--Infeld electrodynamics are stable without omitting $\widetilde{\mathcal{F}}$ dependence.} 

Throughout the paper, we focused only on magnetically charged black holes.  Through the electromagnetic duality of nonlinear electrodynamics \cite{Gibbons:1995cv}, our results are expected to be translated to black holes with an electric charge.
In particular, because Born--Infeld theory is duality invariant, the stability of electrically charged black holes in Born--Infeld electrodynamics follows from that of magnetically charged ones.
 Notice that our analysis cannot apply to black holes with both electric and magnetic charge. Then it is interesting to check the stability of black holes with both magnetic and electric charge.

Our Lagrangian \eqref{eq2-0-1} typically appears as an effective theory for a homogeneous, constant field strength, $\partial_{\mu} F_{\nu\rho} \sim 0$, for example as Euler--Heisenberg action. From the point of view of effective field theory, it is natural to include the derivative interactions like $\partial_{\mu} F_{\nu\rho} \partial^{\mu} F^{\nu\rho}$ for a spherically symmetric case. It is not clear how such interactions affect the stability of black holes.

As an application of our results, it is important to study the quasi-normal modes of gravitational waves and electromagnetic radiations in nonlinear electrodynamics. So far, quasi-normal modes are investigated only without $\widetilde{\mathcal{F}}$ dependence. Thus it is interesting to see effects of $\widetilde{\mathcal{F}}$ dependence on quasi-normal modes. We leave these interesting problems for future work.

\begin{acknowledgments}
D.\,Y. is supported by the JSPS Postdoctoral Fellowships No.201900294. J.\,S. is supported by JSPS KAKENHI Grant Numbers JP17H02894, JP17K18778.
\end{acknowledgments}

\appendix

\section{General nonlinear electrodynamics}
\label{app:General form}

In this appendix, we study the general form of Lagrangian in nonlinear electrodynamics.
Let us define
\begin{align}
\mathcal{X}_{n_{1}, m_{1}, n_{2}, m_{2}, \dots}
\coloneqq
\mathrm{tr}(
F^{n_{1}} \widetilde{F}^{m_{1}} F^{n_{2}} \widetilde{F}^{m_{2}} \cdots
)
\, 
\label{eqA-1}
\end{align}
with non-negative integers $n_{1}, m_{1}, n_{2}, m_{2}, \dots$, where $F$ and $\widetilde{F}$ stand for the matrices with components ${F_{\mu}}^{\nu}$ and ${\widetilde{F}_{\mu}}^{~\nu}$, respectively.
An arbitrary scalar constructed from the field strength and its Hodge dual can be expressed in terms of $\mathcal{X}_{n_{1}, m_{1}, n_{2}, m_{2}, \dots}$.
Our goal in this appendix is to reduce the form of $\mathcal{X}_{n_{1}, m_{1}, n_{2}, m_{2}, \dots}$ to one in terms of the invariants $\mathcal{F} \coloneqq \frac{1}{4} F_{\mu\nu} F^{\mu\nu}$ and $\widetilde{\mathcal{F}} \coloneqq \frac{1}{4} F_{\mu\nu} \widetilde{F}^{\mu\nu} = \frac{1}{8} \epsilon_{\mu\nu\rho\sigma} F^{\mu\nu} F^{\rho\sigma}$.

First, using an identity $\epsilon^{\alpha\beta\gamma\delta} \epsilon_{\mu\nu\rho\sigma} = -4! \delta^{[\alpha}_{\mu} \delta^{\beta}_{\nu} \delta^{\gamma}_{\rho} \delta^{\delta]}_{\sigma}$, we can see that an even number of $\widetilde{F}$ can be expressed in terms of $F$.
Thus we can rewrite $\mathcal{X}_{n_{1}, m_{1}, n_{2}, m_{2}, \dots}$ given by Eq.\,\eqref{eqA-1} as a form including only one or zero $\widetilde{F}$.
Next, we can use the relation $\delta^{[\sigma}_{\mu} \epsilon^{\rho \nu \alpha \beta]} = 0$ to find that
\begin{align}
0 &= F_{\rho\sigma} F_{\alpha\beta} 
\delta^{[\sigma}_{\mu} \epsilon^{\rho \nu \alpha \beta]}
\notag \\
&= \frac{1}{5} F_{\rho\sigma} F_{\alpha\beta} 
\left(
\delta^{\sigma}_{\mu} \epsilon^{\rho\nu\alpha\beta} + \delta^{\rho}_{\mu} \epsilon^{\nu\alpha\beta\sigma} + \delta^{\nu}_{\mu} \epsilon^{\alpha\beta\sigma\rho} +\delta^{\alpha}_{\mu} \epsilon^{\beta\sigma\rho\nu} + \delta^{\beta}_{\mu} \epsilon^{\sigma\rho\nu\alpha}
\right)
\notag \\
&= \frac{1}{5} F_{\rho\sigma} F_{\alpha\beta} \delta_{\mu}^{\nu} \epsilon^{\alpha\beta\sigma\rho}
+ \frac{4}{5}  F_{\rho\sigma} F_{\alpha\beta} \delta_{\mu}^{\sigma} \epsilon^{\rho\nu\alpha\beta}
\notag \\
&= - \frac{8}{5} \delta_{\mu}^{\nu} \widetilde{\mathcal{F}} -\frac{4}{5}  F_{\mu\rho} F_{\alpha\beta} \epsilon^{\rho\nu\alpha\beta} \, .
\label{eqA-2}
\end{align}
In the third line, we use the antisymmetry of $F_{\mu\nu}$ to get together some terms.
Therefore, if the combination of $F\widetilde{F}$ appears in $\mathcal{X}_{n_{1}, m_{1}, n_{2}, m_{2}, \dots}$, we can take it to a form in terms of $\widetilde{\mathcal{F}}$ by using
\begin{align}
{F_{\mu}}^{\rho} {\widetilde{F}_{\rho}}^{~\nu} &= \frac{1}{2} F_{\mu\rho} F_{\alpha\beta} \epsilon^{\rho\nu\alpha\beta}
\notag \\
&= - \delta_{\mu}^{\nu} \widetilde{\mathcal{F}} \, ,
\label{eqA-3}
\end{align} 
where the second line comes from Eq.\,\eqref{eqA-2}.
Thus we can reduce $\mathcal{X}_{n_{1}, m_{1}, n_{2}, m_{2}, \dots}$ to a function of $\widetilde{\mathcal{F}}$ and $\mathrm{tr}(F \cdots F)$.
Finally, the Cayley--Hamilton theorem for $4 \times 4$ matrices states 
\begin{align}
{(F^{4})_{\mu}}^{\nu} &= \mathrm{tr}(F) {(F^{3})_{\mu}}^{\nu} - \frac{1}{2} \left( (\mathrm{tr}(F))^{2} - \mathrm{tr}(F^{2}) \right) {(F^{2})_{\mu}}^{\nu} 
\notag \\
&\quad
+ \frac{1}{6} \left( 
(\mathrm{tr}(F))^{3} - 3 \, \mathrm{tr}(F^{2}) \, \mathrm{tr}(F)
+ 2 \, \mathrm{tr} (F^{3})
 \right) {F_{\mu}}^{\nu} - \det(F) \delta_{\mu}^{\nu} \, ,
\label{eqA-4}
\end{align}
where ${{(F^{4})}_{\mu}}^{\nu}$ means ${F_{\mu}}^{\alpha} {F_{\alpha}}^{\beta} {F_{\beta}}^{\gamma} {F_{\gamma}}^{\nu}$, and so on.
Now we can also use the following relations which hold from the antisymmetry of $F_{\mu\nu}$,
\begin{align}
&\mathrm{tr}(F) = \mathrm{tr}(F^{3}) = 0 \,,
\label{eqA-4-2}\\
&\mathrm{tr}(F^{2}) = {F_{\mu}}^{\nu} {F_{\nu}}^{\mu} = - 4 \mathcal{F} \,,
\label{eqA-4-3}\\
&\det(F) = \frac{1}{g} \det(F_{\mu\nu})
= -\left( \frac{1}{\sqrt{-g}} \mathrm{Pf}(F_{\mu\nu}) \right)^{2}
= - \left( - \frac{1}{8} \epsilon^{\mu\nu\rho\sigma} F_{\mu\nu} F_{\rho\sigma} \right)^{2} = - \widetilde{\mathcal{F}}^{2} \,,
\label{eqA-4-4}
\end{align}
where $\mathrm{Pf}(F_{\mu\nu})$ stands for the Pfaffian of the antisymmetric matrix $F_{\mu\nu}$. 
Note that $F$ has been defined to be a matrix with components ${F_{\mu}}^{\nu}$.
In order to relate the determinant of ${F_{\mu}}^{\nu}$ with that of $F_{\mu\nu}$, additionally we need that of the inverse metric, $g^{-1}$, which appears in the second equality in Eq.\,\eqref{eqA-4-4}.
Note also that we normalize the covariant Levi-Civita tensor as $\epsilon_{0123} = \sqrt{-g}$ throughout this paper, thus $\epsilon^{0123} = - 1/\sqrt{-g}$.
Using these relations, Eq.\,\eqref{eqA-4} becomes
\begin{align}
{(F^{4})_{\mu}}^{\nu} &= \frac{1}{2}\mathrm{tr}(F^{2}) {(F^{2})_{\mu}}^{\nu} 
 - \det(F) \delta_{\mu}^{\nu} 
\notag \\
&= - 2 \mathcal{F} {(F^{2})_{\mu}}^{\nu} +\widetilde{\mathcal{F}}^{2} \delta_{\mu}^{\nu}
\, .
\label{eqA-5}
\end{align}
Since $\mathcal{X}_{n_{1}, m_{1}, n_{2}, m_{2}, \dots}$ is eventually given by the trace of the matrix, we can conclude that it can be reduced to a function of $\mathcal{F}$ and $\widetilde{\mathcal{F}}$.
Therefore, in the context of nonlinear electrodynamics, a general form of Lagrangian is given by a function of the invariants $\mathcal{F}$ and $\widetilde{\mathcal{F}}$: this is what we wanted to show in this appendix.

\section{Decompositions and gauge transformations of perturbations}
\label{app:gauge}

In a study of perturbations on a spherically symmetric background, it is useful to expand the perturbations in terms of spherical harmonics and extract the time and radial dependence of them.
Here we list the decomposition of perturbations explicitly.
First, 10 components of the metric perturbations are expanded as 
\begin{align}
\delta g_{AB} &= \sum_{l,m} \sqrt{\bar{g}_{AA} \bar{g}_{BB}} H_{AB}(l, m; t, r) Y_{lm}(\theta, \phi)
\label{eqB-1} \, ,\\
\delta g_{AI} &= \sum_{l,m} \left[ h^{+}_{A}(l, m; t, r) Y^{+}_{I,lm}(\theta, \phi) + h^{-}_{A}(l, m; t, r) Y^{-}_{I,lm}(\theta, \phi)\right] \,,
\label{eqB-2} \\
\delta g_{IJ} &= \sum_{l,m} \left[ r^{2} K(l, m; t, r) \Omega_{IJ} Y_{lm}(\theta, \phi) 
+
G^{+}(l, m; t, r) Y^{+}_{IJ,lm}(\theta, \phi)
+
G^{-}(l, m; t, r) Y^{-}_{IJ,lm}(\theta, \phi)
\right] \,,
\label{eqB-3}
\end{align}
and 4 components of the gauge potential perturbations are expanded as
\begin{align}
\delta A_{A} &= \sum_{l,m} \delta A_{A}(l,m;t,r) Y_{lm}(\theta,\phi) \,,
\label{eqB-3-2}\\
\delta A_{I} &= \sum_{l,m} 
\left[
\delta A^{+}(l,m;t,r) Y^{+}_{I,lm}(\theta,\phi) + \delta A^{-}(l,m;t,r) Y^{-}_{I,lm}(\theta,\phi)
\right] \,,
\label{eqB-3-3}
\end{align}
where indices $A$ and $B$ take $t$ and $r$, while $I$ and $J$ take $\theta$ and $\phi$.
Here we use $\bar{g}_{AB}$ to represent the components of the metric on the background two-dimensional pseudo-Riemannian manifold where the coordinates $(t,r)$ are given: $\bar{g}_{AB} dx^{A} dx^{B} = - f(r) \, dt^{2} + f^{-1}(r) \, dr^{2}$ in our interest (see Eq.\,\eqref{eq4-3}).
On the other hand, $\Omega_{IJ}$ stand for the components of the metric on the unit two-sphere $S^{2}$: $\Omega_{IJ} dx^{I} dx^{J} = d\theta^{2} + \sin^{2}\theta \, d \phi^{2}$.
The standard (scalar) spherical harmonics are denoted as $Y_{lm}(\theta, \phi)$, where $l$ and $m$ are integers such that $l \geq 0$ and $-l \leq m \leq l$.
In terms of $Y_{lm}(\theta, \phi)$, the vector spherical harmonics defined for $l \geq 1$ are given by
\begin{align}
&Y^{+}_{I,lm}(\theta,\phi) 
= D_{I} Y_{lm}(\theta, \phi) 
= (\partial_{\theta}, \partial_{\phi}) Y_{lm}(\theta, \phi) \, ,
\label{eqB-4}\\
&Y^{-}_{I,lm}(\theta,\phi) 
= {\varepsilon_{I}}^{J} D_{J} Y_{lm}(\theta, \phi)
= \left( \frac{1}{\sin \theta} \partial_{\phi}, - \sin \theta \,\partial_{\theta} \right) Y_{lm}(\theta, \phi) \,,
\label{eqB-5}
\end{align}
where $D_{I}$ and $\varepsilon_{IJ}$ stand for the covariant derivative and the antisymmetric Levi-Civita tensor on $S^{2}$, whose indices are raised or lowered by $\Omega_{IJ}$.
The tensor spherical harmonics defined for $l \geq 2$ are given by
\begin{align}
&Y^{+}_{IJ,lm}(\theta, \phi) 
= D_{I} D_{J} Y_{lm}(\theta,\phi) - {\varepsilon_{I}}^{K} {\varepsilon_{J}}^{L} D_{K} D_{L} Y_{lm}(\theta,\phi)
=
\begin{pmatrix}
W	&X \\
\star	&-\sin^{2} \theta
\end{pmatrix}
Y_{lm}(\theta,\phi) \,,
\label{eqB-6} \\
&Y^{-}_{IJ,lm}(\theta,\phi)
= {\varepsilon_{I}}^{K} D_{K} D_{J} Y_{lm}(\theta,\phi) + {\varepsilon_{J}}^{K} D_{K} D_{I} Y_{lm}(\theta,\phi)
=
\begin{pmatrix}
(1/\sin\theta)X & - \sin \theta \, W\\
\star	& -\sin \theta \, X
\end{pmatrix} Y_{lm}(\theta,\phi) \, ,
\label{eqB-7}
\end{align}
where stars stand for quantities determined by the symmetry of the matrices and some operators are defined by
\begin{align}
W &= \partial_{\theta}^{2} - \cot \theta \, \partial_{\theta} - \frac{1}{\sin^{2}\theta} \partial_{\phi}^{2} \, ,
\label{eqB-8}\\
X &= 2 \partial_{\theta} \partial_{\phi} - 2 \cot \theta \, \partial_{\phi} \, .
\label{eqB-9}
\end{align}
It is convenient to classify the perturbations by their parity since ones which have different parity do not mix in the linear order.
Under parity, note that $Y_{lm}$, $Y^{+}_{I,lm}$ and $Y^{+}_{IJ,lm}$ pick a factor of $(-1)^{l}$, while $Y^{-}_{I,lm}$ and $Y^{-}_{IJ,lm}$ pick $(-1)^{l+1}$.

Then let us consider gauge transformations induced by diffeomorphisms and a $U(1)$ gauge parameter:
\begin{align}
\delta g_{\mu\nu} &\to \delta g_{\mu\nu} + \pounds_{\xi} \bar{g}_{\mu\nu} \,, 
\label{eqB-20}\\
\delta A_{\mu} &\to \delta A_{\mu} + \pounds_{\xi} \bar{A}_{\mu} + \partial_{\mu} \Theta \, ,
\label{eqB-21}
\end{align}
where $\pounds_{\xi}$ represents the Lie derivative along a vector field $\xi_{\mu}$ which produces a set of diffeomorphisms, and $\Theta$ is an arbitrary function corresponding to the additional $U(1)$ gauge symmetry.
The quantities with a bar correspond to the background,
which are given by Eqs.\,\eqref{eq4-3} and \eqref{eq4-4-2} in our case.
A general vector field $\xi_{\mu}$ can be expanded on scalar and vector spherical harmonics as
\begin{align}
\xi_{A} &= \sum_{l,m} \xi_{A}(l,m;t,r) Y_{lm}(\theta,\phi) \,,
\label{eqB-22}\\
\xi_{I} &= \sum_{l,m} 
\left[
\xi^{+}(l,m;t,r) Y^{+}_{I,lm}(\theta,\phi) + \xi^{-}(l,m;t,r) Y^{-}_{I,lm}(\theta,\phi)
\right] \,.
\label{eqB-23}
\end{align}
Given the background as \eqref{eq4-3} and \eqref{eq4-4-2}, we can write down the gauge transformations explicitly.
For the coefficients of metric perturbations, we have
\begin{align}
H_{tt} &\to H_{tt} + \left( \frac{2}{f} \dot{\xi_{t}} - f' \xi_{r} \right) \,,
\label{eqB-33}\\
H_{tr} &\to H_{tr} + \left( \xi_{t}' + \dot{\xi_{r}} - \frac{f'}{f} \xi_{t} \right) \,,
\label{eqB-34}\\
H_{rr} &\to H_{rr} +   \left( f' \xi_{r} + 2 f\xi_{r}'  \right) \,,
\label{eqB-35}\\
h_{t}^{+} &\to h_{t}^{+} + \left( \xi_{t} + \dot{\xi}^{+} \right) \,,
\label{eqB-36}\\
h_{t}^{-} &\to h_{t}^{-} + \dot{\xi}^{-} \,,
\label{eqB-30}\\
h_{r}^{+} &\to h_{r}^{+} + \left( \xi_{r} + \xi^{+\prime} - \frac{2}{r}\xi^{+} \right) \,,
\label{eqB-37}\\
h_{r}^{-} &\to h_{r}^{-} + \left( \xi^{-\prime} - \frac{2}{r}\xi^{-} \right) \,,
\label{eqB-31}\\
K &\to K + \left( \frac{2 f }{r}\xi_{r} - \frac{l(l+1)}{r^{2}} \xi^{+} \right) \,,
\label{eqB-39}\\
G^{+} &\to G^{+} + \xi^{+} \,,
\label{eqB-38}\\
G^{-} &\to G^{-} + \xi^{-} \,,
\label{eqB-32}
\end{align}
where a dot and a prime denote the derivative with respect to $t$ and $r$, respectively.
For $l \geq 2$, we can choose the 4 degrees of freedom of gauge functions so that $h^{+}_{t} = h^{+}_{r} = G^{+} = G^{-} = 0$.
After that and renaming
\begin{align}
H_{0} &= H_{tt} \,,
\label{eqB-40}\\
H_{1} &= H_{tr} = H_{rt} \,,
\label{eqB-41}\\
H_{2} &= H_{rr} \,,
\label{eqB-42}\\
h_{0} &= - h^{-}_{t} \,,
\label{eqB-43}\\
h_{1} &= - h^{-}_{r} \, ,
\label{eqB-44}
\end{align}
which follows the notation in Ref.\,\cite{Regge:1957td}, we are left with Eqs.\,\eqref{eq4-7} and \eqref{eq4-8}.
For $l = 1$, $G^{+}$ and $G^{-}$ are not defined, thus we can set $h^{+}_{t} = h^{-}_{t}(= - h_{0}) = h^{+}_{r} = K = 0$ by choosing four degrees of freedom of gauge functions appropriately. 
For $l = 0$, the functions $h^{+}_{t}$, $h^{-}_{t}$, $h^{+}_{r}$, $h^{-}_{r}$, $G^{+}$ and $G^{-}$ are not defined and we have two degrees of freedom of gauge denoted by $\xi_{t}$ and $\xi_{r}$.
Then we can choose a gauge fixing such that $H_{tr}(=H_{1}) = K = 0$.

Let us turn to gauge transformations of  electromagnetic perturbations.
When we expand a $U(1)$ gauge parameter $\Theta$ in Eq.\,\eqref{eqB-21} as
\begin{align}
\Theta (t, r, \theta, \phi)
= \sum_{l,m} \left[ 
\Theta(t,r) Y_{lm}(\theta, \phi) - \frac{q(\pm1 - \cos \theta)}{r^{2} \sin \theta} 
\left( \xi^{+} \frac{1}{\sin \theta} \partial_{\phi} Y_{lm} - \xi^{-} \partial_{\theta} Y_{lm} \right)
\right] \,,
\label{eqB-50}
\end{align}
where $\Theta (t, r)$ is an arbitrary function of $(t, r)$,
the gauge transformations are induced as follows:
\begin{align}
\delta A_{t} &\to \delta A_{t} + \dot{\Theta} \,,
\label{eqB-51}\\
\delta A_{r} &\to \delta A_{r} + \Theta' \,,
\label{eqB-52}\\
\delta A^{+} &\to \delta A^{+} + \left( \Theta + \frac{q }{r^{2}} \xi^{-} \right) \,,
\label{eqB-53}\\
\delta A^{-} &\to \delta A^{-} - \frac{q}{r^{2}} \xi^{+}
\label{eqB-54} \,.
\end{align}
For $l \geq 1$, we can use the degree of freedom of gauge parameter $\Theta$ to set $\delta A^{+} = 0$.
After that and renaming 
\begin{align}
\frac{f^{-}_{23}}{l(l+1)} &= \delta A^{-} \,,
\label{eqB-55}\\
f^{+}_{02} &= - \, \delta A_{t} \,,
\label{eqB-56}\\
f^{+}_{12} &= - \, \delta A_{r} \,,
\label{eqB-57}
\end{align}
we have the expressions in the main text, Eqs.\,\eqref{eq4-12}--\eqref{eq4-15}, \eqref{eq4-18}--\eqref{eq4-21}.
For $l = 0$, the functions $\delta A^{+}$ and $\delta A^{-}$ are not defined, thus we can be left only with $\delta A_{r}$ (or equivalently $f^{+}_{12}$) by using the degree of freedom of gauge to remove $\delta A_{t}$ (or $f^{+}_{02}$).

Before the gauge fixing, we can see the following combinations are gauge invariant:
\begin{align}
\widehat{h}^{-}_{t} &\coloneqq h_{t}^{-} - \dot{G}^{-} \,,
\label{eqB-60}\\
\widehat{h}_{r}^{-} &\coloneqq h_{r}^{-} - G^{-\prime} + \frac{2}{r} G^{-} \,,
\label{eqB-61}\\
\widehat{H}_{tt} &\coloneqq H_{tt} - \frac{2}{f} \left( \dot{h}^{+}_{t} - \ddot{G}^{+} \right) + f' \left( h^{+}_{r} - G^{+\prime} + \frac{2}{r} G^{+} \right) \,,
\label{eqB-62}\\
\widehat{H}_{tr} &\coloneqq H_{tr} - \left( h^{+}_{t} - \dot{G}^{+} \right)' - \left( \dot{h}^{+}_{r} - \dot{G}^{+\prime} + \frac{2}{r} \dot{G}^{+} \right) + \frac{f'}{f} \left( h^{+}_{t} - \dot{G}^{+} \right) \,,
\label{eqB-63}\\
\widehat{H}_{rr} &\coloneqq H_{rr} - f' \left( h^{+}_{r} - G^{+\prime} + \frac{2}{r} G^{+} \right) - 2 f \left( h^{+}_{r} - G^{+\prime} + \frac{2}{r} G^{+} \right)' \,,
\label{eqB-64}\\
\widehat{K} &\coloneqq K - \frac{2f}{r} \left( h^{+}_{r} - G^{+\prime} + \frac{2}{r} G^{+} \right) + \frac{l(l+1)}{r^{2}} G^{+} \,,
\label{eqB-65}\\
\delta \widehat{A}^{-} &\coloneqq \delta A^{-} + \frac{q}{r^{2}} G^{+} \,,
\label{eqB-66}\\
\delta \widehat{A}_{t} &\coloneqq \delta A_{t} - \left( \delta \dot{A}^{+} - \frac{q}{r^{2}} \dot{G}^{-} \right) \,,
\label{eqB-67}\\
\delta \widehat{A}_{r} &\coloneqq \delta A_{r} - \left( \delta {A}^{+} - \frac{q}{r^{2}} {G}^{-} \right)' \,.
\label{eqB-68}
\end{align}


\bibliography{biblio}

\end{document}